\documentclass[12pt]{article}

\usepackage{template}
\usepackage{soul}
\usepackage[utf8]{inputenc} 
\usepackage[T1]{fontenc}    
\usepackage{booktabs}       
\usepackage{amsfonts}       
\usepackage{nicefrac}       
\usepackage{microtype}      
\usepackage{lipsum}
\usepackage{fancyhdr}       
\usepackage{graphicx}       
\usepackage{makecell}
\usepackage{tablefootnote}
\graphicspath{{media/}}     
\usepackage{authblk}
\usepackage{float}
\usepackage[section]{placeins}
\usepackage{lineno}
\usepackage{bm}
\usepackage{amsmath}
\usepackage{geometry}
\usepackage{xr}

\usepackage[backend=bibtex, maxnames=10, minnames=1]{biblatex}
\usepackage{url}            
\PassOptionsToPackage{hyphens}{url}\usepackage{hyperref}
\usepackage{xurl}

\newcommand\thanksfootnote[1]{%
  \begingroup
  \renewcommand\thefootnote{}\footnote{#1}%
  \addtocounter{footnote}{-1}%
  \endgroup
}

\title{Fast, Scale-Adaptive, and Uncertainty-Aware Downscaling of Earth System Model Fields with Generative Machine Learning}

\author[1,2]{Philipp Hess}
\author[1]{Michael Aich}
\author[3]{Baoxiang Pan}
\author[1,2,4]{Niklas Boers}

\affil[1]{\small Technical University Munich, Munich, Germany; School of Engineering \& Design, Earth System Modelling}
\affil[2]{Potsdam Institute for Climate Impact Research, Potsdam, Germany}
\affil[3]{Institute of Atmospheric Physics, Chinese Academy of Sciences, Beijing, China}
\affil[4]{Global Systems Institute and Department of Mathematics, University of Exeter, Exeter, UK}

\begin{document}

\maketitle

\begin{abstract}
    Accurate and high-resolution Earth system model (ESM) simulations are essential to assess the ecological and socio-economic impacts of anthropogenic climate change, but are computationally too expensive to be run at sufficiently high spatial resolution. Recent machine learning approaches have shown promising results in downscaling ESM simulations, outperforming state-of-the-art statistical approaches. However, existing methods require computationally costly retraining for each ESM and extrapolate poorly to climates unseen during training. We address these shortcomings by learning a consistency model (CM) that efficiently and accurately downscales arbitrary ESM simulations without retraining in a zero-shot manner.
    Our approach yields probabilistic downscaled fields at a resolution only limited by the observational reference data. 
    We show that the CM outperforms state-of-the-art diffusion models at a fraction of computational cost while maintaining high controllability on the downscaling task. Further, our method generalizes to climate states unseen during training without explicitly formulated physical constraints.
\end{abstract}

\thanksfootnote{Contact: Philipp Hess, philipp.hess@tum.de}

\newgeometry{textwidth=5.5in}
\section{Introduction}

Accurate and high-resolution climate simulations are of crucial importance in projecting the climatic, hydrological, ecological, and socioeconomic impacts of anthropogenic climate change. 
Precipitation is one of the most important climate variables, with particularly large impacts, e.g., on vegetation and crop yields, infrastructure, or the economy \cite{kotz_effect_2022}. However, it is also the variable that is arguably most difficult to model and predict, especially extreme precipitation events, resulting from a complex interaction of processes over a large range of spatial and temporal scales that cannot be fully resolved.  

Numerical Earth system models (ESMs) are our main tool to project the future evolution of precipitation and its extremes. 
However, they exhibit biases and have much coarser spatial resolution, on the order of 10-100km, than needed for reliable assessments of the impacts of climate change \cite{schneider_climate_2017}. Therefore, ESM projections have to be downscaled to higher resolution, for which several machine learning-based approaches have recently been proposed. Due to the chaotic nature of geophysical fluid dynamics, the trajectory of climate simulations will not match historical observations, requiring approaches suitable for unpaired samples to learn such tasks. Recently, methods from generative deep learning have shown promising results in downscaling or correcting spatial patterns of ESM simulations. Normalizing flows (NFs) \cite{dinh_nice_2015} and generative adversarial networks (GANs) \cite{goodfellow_generative_2014} can perform these tasks efficiently in a single step \cite{groenke_climalign_2020, pan_learning_2021, francois_adjusting_2021, harris_generative_2022, hess_physically_2022, hess_deep_2023}. However, NFs often exhibit lower quality, e.g., less sharp and detailed generated images, while GANs can suffer from training instabilities and problems such as mode collapse \cite{arjovsky_towards_2017}.
Moreover, these approaches require computationally expensive retraining of the neural networks for each specific ESM simulation to be processed. This makes downscaling large ESM ensembles, as needed in impact assessments, prohibitively costly and time-consuming.

Diffusion-based generative models have demonstrated superior performance over NFs and GANs on classical image generation tasks \cite{dhariwal_diffusion_2021, song_score-based_2021}. Crucially, iteratively solving the reversed diffusion equation allows for strong control over the image sampling process. 

This enables foundation models for computer vision and image processing, which are only trained to generate target dataset samples from noise and which can later be repurposed for different downstream tasks, e.g., inpainting, classification, colorization, or generating realistic images based on a given ``stroke sketch'' guide as with SDEdit \cite{clark_text--image_2023, sauer_adversarial_2024, meng_sdedit_2022}.

So far, such diffusion model-based approaches have only been applied to downscale idealized fluid dynamics \cite{bischoff_unpaired_2024, wan_debias_2023} in Earth system science-related tasks. Both studies use stochastic differential equation (SDE) based diffusion models and achieve remarkable performance. The iterative integration of the SDE, however, implies that the generative network needs to be evaluated up to 1000 times in order to downscale a single simulated field. This makes such methods unsuitable for processing large simulation datasets, e.g., at high temporal resolution or over long periods, as would be needed in the context of climate change projections and impact assessments.

Much effort has been taken to improve the sampling speed of diffusion models. Using ordinary differential equations instead of SDEs can reduce the number of integration steps to around 10-50 \cite{karras_elucidating_2022, esser_scaling_2024}. Distillation techniques can also be applied to improve the sampling speed to a single step \cite{luhman_knowledge_2021, zheng_fast_2024}. However, they require additional training costs and hyperparameter choices. 
In general, a trade-off between the number of generative steps and the resulting image quality scores has been found \cite{song_score-based_2021, esser_scaling_2024}.
Recently, consistency models (CMs) have been proposed that can be directly trained as single-step generative models without sacrifices in the controllability of the sampling \cite{song_consistency_2023}. CMs have been shown to outperform distilled single-step diffusion models on common image benchmarks \cite{song_consistency_2023}. 

In this work, we tackle the shortcomings of standard diffusion models by using CMs \cite{song_consistency_2023} to downscale global precipitation simulations from three different ESMs in a single step. In summary, we aim to establish a generative model for downscaling ESM fields that fulfills the following conditions:
\begin{itemize}
    \item The CM model is trained on the target dataset only, without conditioning on the ESM. This makes the method applicable to any ESM without requiring computationally expensive retraining.
    
    \item The training minimizes a regression loss that is more stable than the adversarial training in previous methods \cite{pan_learning_2021, francois_adjusting_2021, hess_physically_2022, hess_deep_2023}.
    
    \item The generative downscaling is controllable at the inference stage after training, allowing the choice of a characteristic spatial scale up to which unbiased spatial patterns are preserved in the ESM.
    
    \item The model only requires a single network evaluation instead of the many iterative evaluations in diffusion model-based downscaling approaches \cite{bischoff_unpaired_2024, wan_debias_2023}.
    
    \item The model allows to generate a large number of downscaled realizations for a single ESM field (i.e., a one-to-many mapping), enabling probabilistic downscaling and a quantification of the sampling spread.
    
    \item The method does not require specifically formulated physical constraints \cite{harder_hard-constrained_2023} in order to preserve trends \cite{hess_physically_2022}.
\end{itemize}

\begin{figure}[H]
    \centering
    \includegraphics[width=0.95\textwidth]{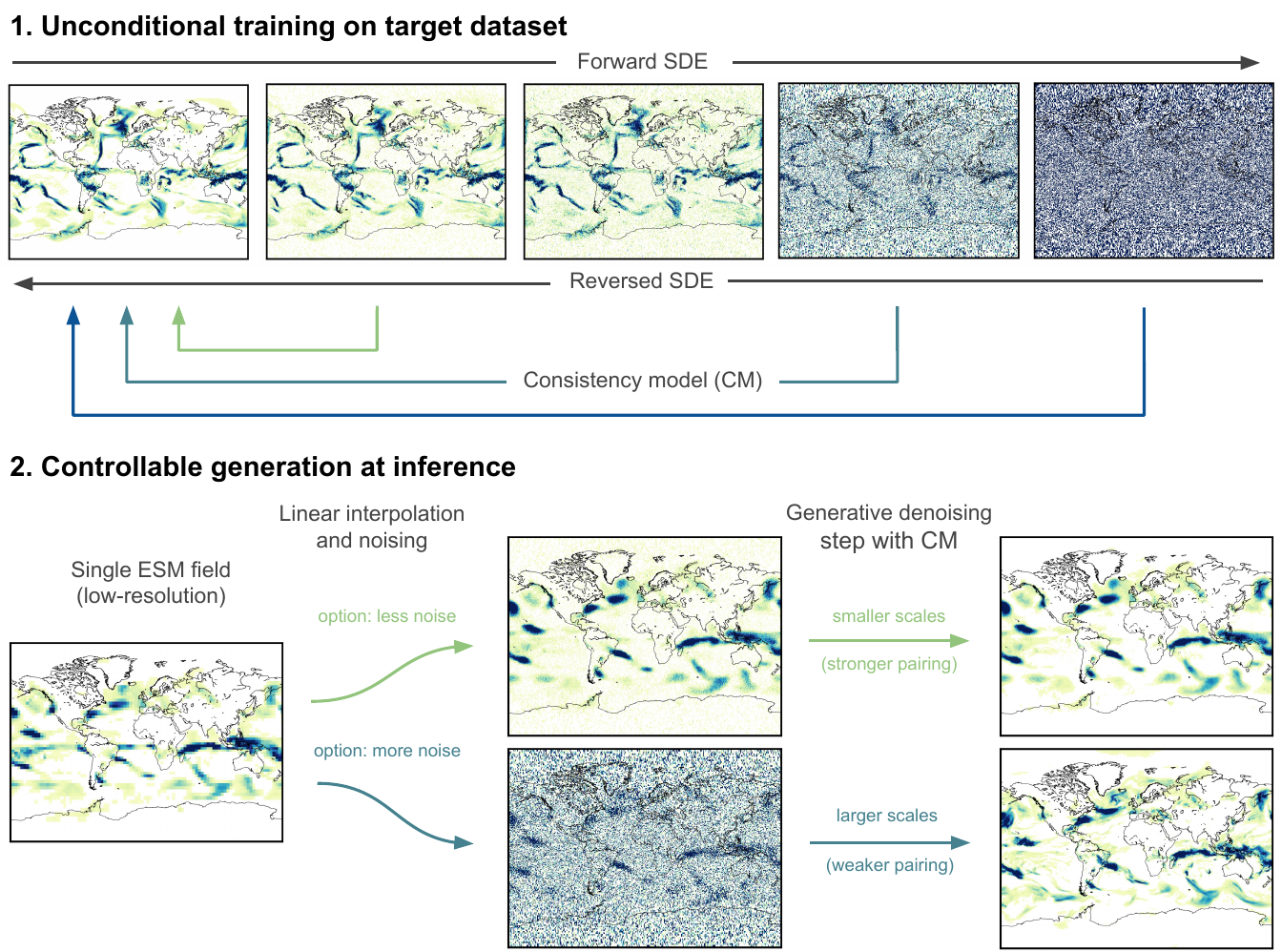}
    \caption{\small{Sketch of the consistency model for downscaling of Earth system model fields. (Upper panel) Unconditional training of the score-based diffusion and consistency models (CM) that learn to reverse a forward diffusion process. While the stochastic differential equation of the diffusion model requires an iterative integration over many steps, the CM only takes a single step to generate a global precipitation field from noise. (Lower panel) The unconditionally trained consistency model is used to downscale (upsample) a low-resolution ESM precipitation field to a four times higher resolution. By adding noise of a chosen variance to the ESM field, the spatial scale to be preserved in the ESM can be controlled: small noise variance implies a close pairing to the original ESM field with only small changes; a larger variance will result in changes at larger spatial scales and in weaker pairing to the ESM field.}}
    \label{fig:method_overview}
\end{figure}

\section{Results}

We evaluate our CM-based downscaling method against the SDE bridge from \cite{bischoff_unpaired_2024} over the test set data.  We investigate the performance of the downscaling, the ability to correct distributional biases in the ESM, the sample spread given a chosen spatial scale, and the preservation of trends in future climate scenarios. 

\subsection{Downscaling Spatial Fields}

\begin{figure}[H]
    \centering
    \includegraphics[width=0.8\textwidth]{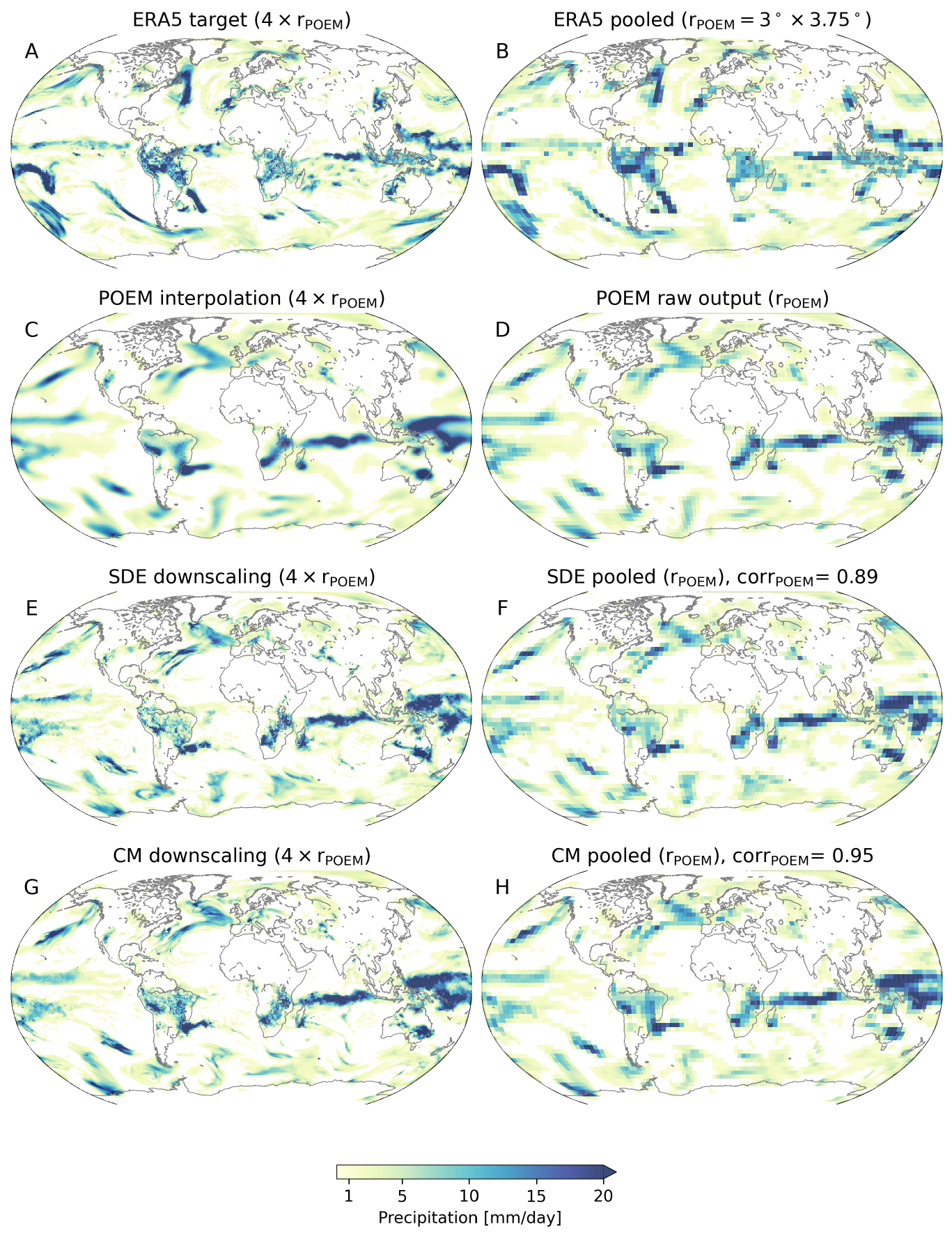}
    \caption{\small{Qualitative comparison of single-day precipitation fields. (A) Daily precipitation from the ERA5 target dataset was used for training the generative models. (B) Same as (A) but at four times lower resolution for comparisons. (C) A precipitation field from a historical run of the POEM ESM interpolated to the target resolution and (D) on its native resolution of $3^\circ \times 3.75^\circ$. The POEM fields are unpaired with the ERA5 field from the same date or any other ERA5 field. Downscaled field from POEM (D) with the SDE bridge method (D)$\rightarrow$(E). (F) An upscaled (average pooled) representation of (E) for comparison with the original POEM field is shown in D and the Pearson correlation between the two. 
    Downscaling POEM with the CM-based method (D)$\rightarrow$(G),
    and the respective pooled field (H). Note that the CM downscaling yields a higher correlation, and hence better consistency of the large-scale features, than the SDE method.}}
    \label{fig:global_fields}
\end{figure}

For a qualitative comparison, we show single precipitation fields in Fig.~\ref{fig:global_fields}. Both generative downscaling methods, based on the SDE bridge (Fig.~\ref{fig:global_fields}E) and CM (Fig.~\ref{fig:global_fields}G), respectively, are able to produce high-resolution precipitation fields that are visually indistinguishable from the unpaired ERA5 field (Fig.~\ref{fig:global_fields}A). A similar performance can be seen when downscaling simulations from the state-of-the-art GFDL-ESM4, the more lightweight SpeedyWeather.jl general circulation model, as well as when applying our method to initially upscaled ERA5 data as a proof of concept (Fig.~S3, S6).

When upscaled back to the native ESM resolution using a $4 \times 4$-kernel average pooling, an accurate representation of the low-resolution ESM simulation field is apparent. As indicated in Fig.~\ref{fig:global_fields}F and Fig.~\ref{fig:global_fields}H for the case of the POEM model, a high Pearson correlation of $0.89$ and $0.95$ for the SDE and CM methods, respectively, is maintained between upscaled corrected and native fields. We provide correlation statistics for the entire test set in Table \ref{tab:correlation}. Besides the average pooled fields, we also compare the downscaled and linearly interpolated POEM simulation on the high-resolution grid by applying a low-pass filter with a cut-off frequency set to $k^*=0.0667$ on the downscaled fields before computing the correlation. In this way, we test the preservation of the large-scale patterns in the ESM. More specifically, we measure how similar the downscaled fields are to the original coarse fields when upscaling them back to the same resolution. The SDE bridge achieves a mean correlation of $0.918$ and $0.916$ for the average pooled and low-pass filtered fields, respectively. Our CM-based method achieves even higher correlation values of $0.954$ and $0.941$ for both measures. The CM-based downscaling also shows higher correlations when downscaling ERA5 (see Table S2).

We estimate the average time it takes to produce a single sample with the SDE and CM methods on a NVIDIA V100 32GB GPU. The average is taken over $100$ samples, and we set the number of SDE integration steps to $500$ as in \cite{bischoff_unpaired_2024}, which is lower than the typical $1000-2000$ steps \cite{dhariwal_diffusion_2021, song_score-based_2021}. The SDE takes on average $39.4$ seconds, while the CM samples much more efficiently, taking only $0.1$ seconds, scaling linearly with the number of samples.

\begin{figure}[H]
    \centering
    \includegraphics[width=0.8\textwidth]{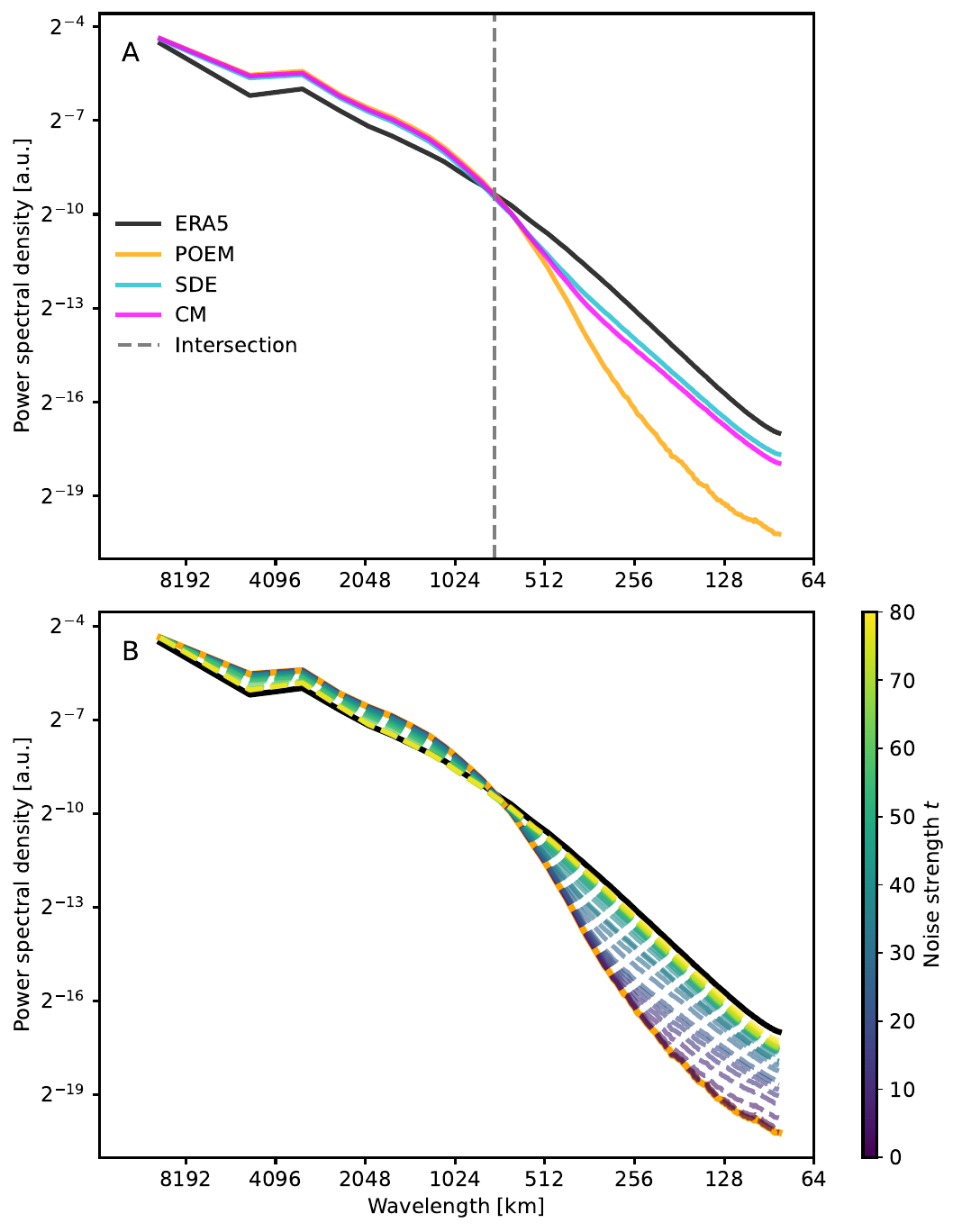}
    \caption{\small{Mean spatial power spectral densities (PSDs) of global precipitation fields. (A) Comparison of the PSDs for the target ERA5 reanalysis data (black), the POEM simulations interpolated to the same high-resolution grid (orange), the SDE bridge (cyan), and the CM downscaling (magenta). The vertical dashed lines mark the spatial scale at which the PSDs of POEM and ERA5 intersect and are thus a natural choice for the wavenumber $k^{\ast}$ up to which to correct, which in turn determines $t^{\ast}$, i.e. the noising strength in the diffusion models (see Eq.~\ref{eq:stop_time} in the Methods). (B) CM downscaling (dashed lines) applied to be consistent with different spatial scales as a function of the noising strength $t$ over the entire range $[t_{\mathrm{min}}, t_{\mathrm{max}}]$. Noising small scales implies nearly reproducing the POEM simulations, while noising larger scales corresponds to a weaker pairing to the ESM (see Fig.~\ref{fig:method_overview}).}}
    \label{fig:psds}
\end{figure}

We analyze the downscaling performance of the CM and SDE bridge approaches quantitatively using power spectral densities (PSDs) as e.g. in \cite{ravuri_skilful_2021, hess_deep_2023}. 
The interpolated ESM fields under-represent variability at small spatial scales, since these are not present in the low-resolution simulation. This implies an underestimation of spatial intermittency, i.e., overly smooth precipitation patterns, which is highly problematic for impact assessment. The generative downscaling methods based on SDEs and CM perform very well in increasing spatial resolution and greatly improve the spatial intermittency at the smaller spatial scales (Fig.~\ref{fig:psds}A). The PSD improvements are found to be consistent when applying the CM downscaling to the GFDL-ESM4 and SpeedyWeather.jl simulations (Fig.~S4).

We also investigate the change in PSDs as a function of the noise variance schedule time $t$ in Fig.~\ref{fig:psds}B, which is directly related the spatial scale up to which patterns are corrected (see Methods). For minimal noise, i.e., $t^* = t_{\mathrm{min}}$, the CM model reproduces the PSD of the ESM, as expected from Eq.~\ref{eq:cm_coeff} since there are no changes made to the ESM field. For maximal variance with $t^* = t_{\mathrm{max}}$, the PSD closely matches the ERA5 reanalysis ground truth. For any $ t_{\mathrm{min}} < t^* < t_{\mathrm{max}}$, we find a trade-off between the two extreme cases that match the PSD above and below the intersection (dotted grey line) to a certain degree, depending on the spatial scale preserved in the ESM.

To test the temporal consistency of the downscaled fields, we report temporal autocorrelation values computed for each grid cell as global averages in Fig.~S8. We find that the generative and stochastic SDE and CM methods result in a more accurate temporal correlation than the deterministic approach of quantile mapping and bilinear interpolation.

\begin{table}[H]
    \small
    \centering
        \caption{\small{Summary statistics comparing our CM approach and the SDE bridge as a benchmark. Correlations are computed (2nd column) on the native POEM grid ($r = 3^\circ \times 3.75^\circ)$ by applying a $4 \times 4$ average pooling to the downscaled fields and (3rd column) on the downscaled grid by applying a low-pass filter to the high-resolution downscaled fields with a cut-off frequency that is consistent with the chosen spatial scale in POEM to be preserved. (4th column) The global long-term mean absolute error with respect to ERA5 and (5th column) the respective error reduction compared to the POEM ESM bias. (6th column) The global absolute error in the 95$^{\mathrm{th}}$ precipitation percentile and (7th column) the error reduction with respect to the POEM ESM bias. An estimation of the mean sampling time is also reported (8th column).}}
         \tablefootnote{The best performance is given in bold font.}
        \small{
    \begin{tabular}{r c c c c c c c}
        \toprule
        Model & \makecell{Correlation \\ (pooled)}  & \makecell{Correlation \\ (low-pass) } &  \makecell{Mean \\ error}  &  \%  & \makecell{95$^{\mathrm{th}}$ precentile\\ error} & \% & \makecell{Sample \\ time [s] }\\
        \hline
        SDE & 0.918          & 0.916            &  \textbf{0.214} & \textbf{72.51} & 1.106 & 68.15 & 39.4          \\
        CM  & \textbf{0.954} & \textbf{0.941}   &  0.217 & 72.08 & \textbf{1.080} & \textbf{68.92} & \textbf{0.1}  \\
        \bottomrule
        \label{tab:correlation}
    \end{tabular}       
    }
\end{table}

\subsection{Bias Correction}
\label{sec:results_biases}

We compare the ability to correct biases in the ESM with histograms of relative frequency and latitude-profiles of mean precipitation to investigate the reduction of known biases such as a double-ITCZ \cite{tian_double-itcz_2020}, following the evaluation methodology in \cite{hess_physically_2022, hess_deep_2023}.

When applied to the ESM simulations without preprocessing via quantile delta mapping (QDM) \cite{cannon_bias_2015} , the ability of the CM to correct biases naturally depends on the chosen spatial consistency scale (Fig.~S9). Selecting the smallest scale reproduces the ESM without any changes, hence inheriting its biases. Choosing the largest possible scales generates samples with statistics very close to the target dataset. However, the fields become more and more unpaired to the ESM at such high noise levels (see Fig.~\ref{fig:method_overview}). A scale between these two extremes will correct for biases to a varying degree, depending on the chosen correction scale.

\begin{figure}[H]
    \centering
    \includegraphics[width=1.0\textwidth]{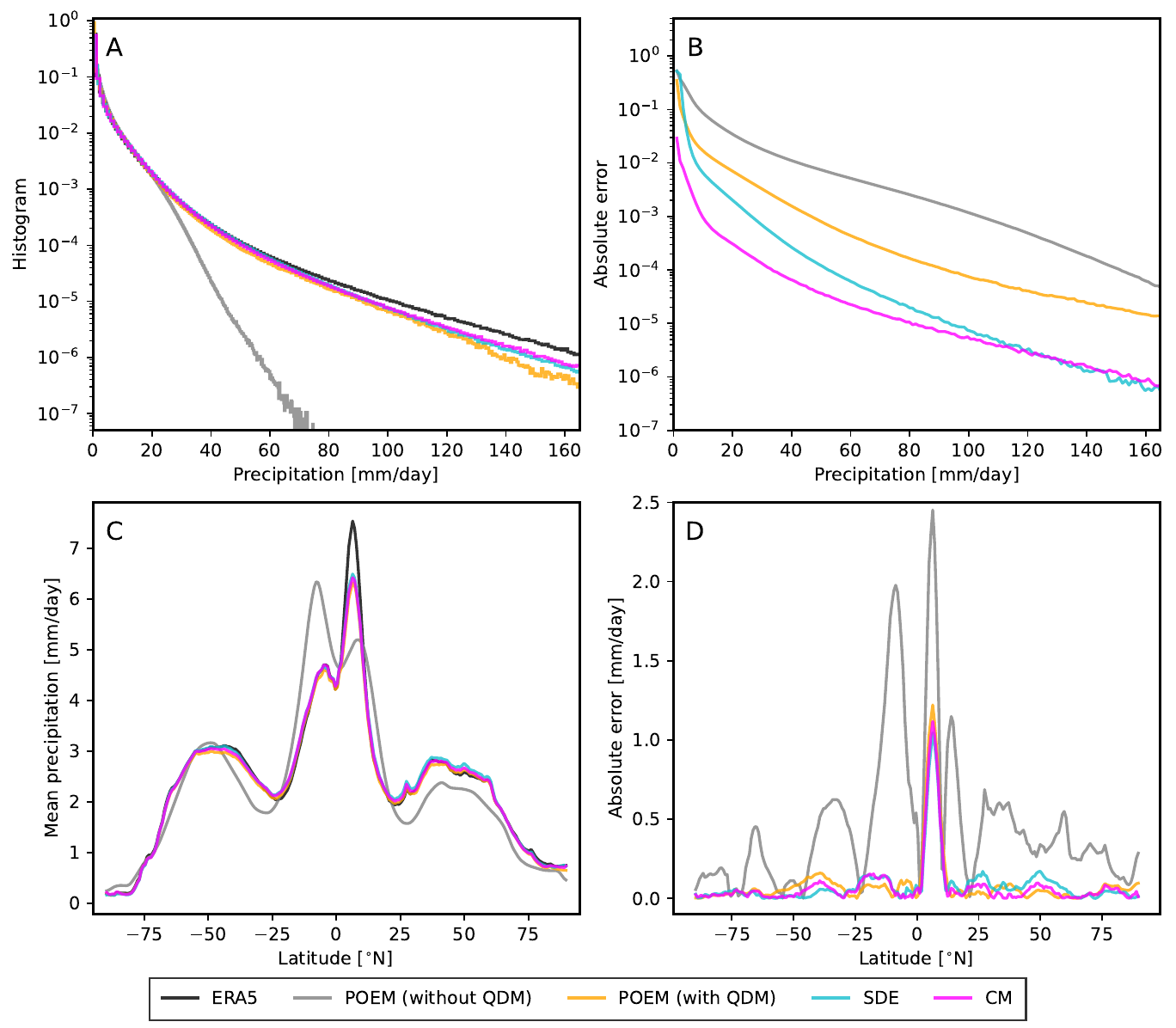}
    \caption{\small{Comparison of global histograms and longitudinal mean precipitation. (A) Global histograms of relative precipitation frequency for the ERA5 reanalysis data (black), POEM simulations without applying the QDM-preprocessing (grey), POEM simulations with QDM (orange), the SDE bridge (cyan), and the CM (magenta).
    (B) Absolute errors of the histograms in (A) with respect to the ERA5 ground truth. (C) Precipitation averaged over time and longitudes for the same data as in (A). (D) Absolute errors of the latitude profile in (C). 
    Both the SDE and the CM downscaling method are able to further improve upon the QDM-preprocessing in terms of bias correction, most notably for extreme precipitation.}}
    \label{fig:biases}
\end{figure}

In terms of relative frequency histograms, the ESM simulations (without QDM preprocessing) exhibit a very strong under-representation of the right tail of the distribution, i.e., of the extremes (Fig.~\ref{fig:biases}A and Fig.~\ref{fig:biases}B). This misrepresentation of extremes is a key problem with existing state-of-the-art ESMs and makes future projections of extreme events and their impacts, as well as related detection and attribution of extremes, highly uncertain. 

Applying QDM to POEM strongly improves the frequency distributions as expected. Downscaling the ESM with the SDE further improves the global histograms by an order magnitude, particularly for the extremes. Our CM-based method shows the overall largest bias reduction in the global histograms (Fig.~\ref{fig:biases}B).
 When applied to the different ESMs and the upscaled ERA5 data, our method shows a consistent bias correction skill (Fig.~S5). 

We further compute the error in the 95$^{\mathrm{th}}$ percentile of the local precipitation histogram for each grid cell and aggregate the absolute value globally (Tab.~\ref{tab:correlation}). The SDE method shows an error of $1.106$ mm/day, reducing the error of the POEM ESM by $68.15\%$. The CM method performs slightly better with an error of $1.08$ mm/day and a respective error reduction of $68.92\%$ of the ESM. When downscaling the initially upsampled ERA5 data, the CM again shows better performance, with an error of $0.725$ mm/day compared to the SDE with an error of 0.868 mm/day (Tab.~S2).

The POEM model exhibits a strong double-ITCZ bias that is common among state-of-the-art ESMs \cite{tian_double-itcz_2020} (Fig.\ref{fig:biases}C). As expected, QDM is able to remove most of the biases, though slightly underestimating the peak north of the equator in the ITCZ. The downscaling methods based on the SDE bridge and CM show a similar absolute error for these latitude profiles as when only applying QDM alone (Fig.\ref{fig:biases}D).  

We report the absolute value of the grid cell-wise error in the long-term mean, again averaged globally, for both downscaling methods (Tab.~\ref{tab:correlation}). The SDE downscaling results in an error of $0.214$ mm/day, reducing the error in the POEM model by $72.51 \%$, and performing slightly better than our CM method, which exhibits an error of $0.217$ mm/day and a respective error reduction of $72.08\%$ when applied to POEM.  Evaluated on the ERA5 data, we find that the CM method performs slightly better than the SDE in terms of mean absolute error and RMSE (Tab.~S2).

\subsection{Quantifying the Sampling Spread}

Our generative CM-based downscaling is stochastic, with a one-to-many mapping of a single ESM field to many possible downscaled realizations. It thus naturally yields a probabilistic downscaling, suitable to estimate the associated uncertainties. By selecting a given spatial scale in the ESM to be preserved by the downscaling method, one automatically chooses a related degree of freedom to generate patterns on smaller scales. Given that our CM method is very efficient at inference, we can generate a large ensemble of high-resolution fields that are consistent with the low-resolution ESM input, and compute statistics such as the sampling spread, which can be interpreted as a measure of the inherent uncertainty of the downscaling task. 

\begin{figure}[H]
    \centering
    \includegraphics[width=0.75\textwidth]{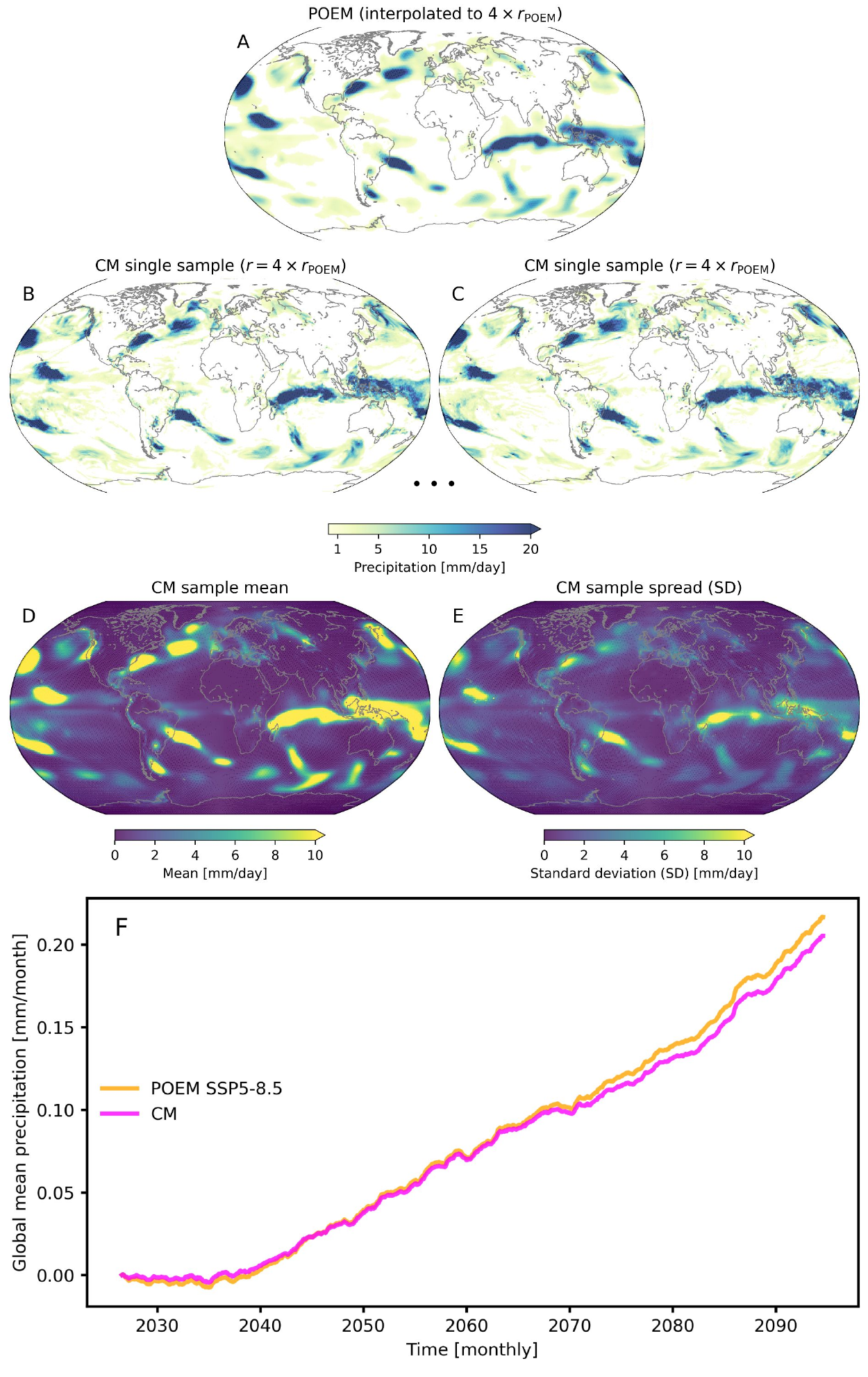}
    \caption{\small{Sampling spread and generalization to unseen climates of the generative probabilistic downscaling process. (A) The ESM field interpolated to the target resolution. (B) and (C) show two different exemplary samples generated by the CM downscaling, preserving large-scale patterns and generating new patterns on smaller scales. (D) The ensemble mean of $10^3$ samples with the standard deviation is shown in (E). (F) Three-year rolling global mean normalized to the reference year 2020 of the very high emission scenario SSP5-8.5. The ESM (orange) shows an increase in global mean precipitation over the emission scenario, in line with the thermodynamic Clausius-Clapeyron relation \cite{traxl_role_2021}. The CM downscaling (magenta) is able to preserve the trend with a high degree of accuracy, notably without the addition of any physical constraints in the CM network.}}
    \label{fig:uncertaint_quantification}
\end{figure}

We compute an ensemble of $10^3$ downscaled fields from a single ESM precipitation field (Fig.~\ref{fig:uncertaint_quantification}A) and evaluate the mean and standard deviation (Fig.~\ref{fig:uncertaint_quantification}D and Fig.~\ref{fig:uncertaint_quantification}E). The ensemble mean shows close similarity to the ESM simulation interpolated to the same high-resolution grid. The sample spread shows patterns similar to the mean, although with a smaller magnitude.

We evaluate an ensemble of 100 downscaled realizations of coarse ERA5 fields using the continuous ranked probability score (CRPS) \cite{hersbach_decomposition_2000}, for three different noise scales, corresponding to $t=t_{\text{min}}$, $t=t^*$, and $t=t_{\text{max}}$ (Fig.~S7). Further, the CRPS is evaluated as a skill score (CRPSS) relative to the baseline ensemble of 100 random high-resolution ERA5 fields. We find that the noise strength enables calibration of the ensemble spread, which is sharp for small noise scales and increasing broad for larger noise scales. The intermediate noise scale $t=t^*$ shows the lowest CRPS and maximum noise leads to a similar CRPSS to the baseline climatology of drawing random fields. For definitions of the CRPS and CRPSS see section 8 in the SI.

\subsection{Future Projections}

The efficient nature of our CM approach enables the downscaling of long climate simulations over a century, such as prescribed with the SPP5-8.5 high-emission scenario from the Coupled Model Intercomparison Project Phase 6 (CMIP6). We find that the downscaling accurately preserves the global precipitation content of the ESM and preserves the non-linear trends, as expected from the Clausius-Clapeyron relation \cite{traxl_role_2021} (Fig.~\ref{fig:uncertaint_quantification}F).
The network thus generalizes to the out-of-sample predictions of increasing global mean precipitation without the need for hard-coding auxiliary physical constraints in the network as, for example, done in \cite{beucler_enforcing_2021, harder_hard-constrained_2023, hess_physically_2022}. However, our method can, in principle, be extended to such constraints to preserve trends exactly. 
We note that the ability to preserve trends is expected to be related to the noise variance used for conditioning the downscaling, as maximum noise corresponds to sampling from the learned historical ERA5 distribution, and no noise does not alter the ESM simulation in any way.

\section{Discussion}

We introduced a generative machine learning approach for efficient, scale-adaptive, and probabilistic downscaling of ESM precipitation fields. Our approach is based on consistency models (CMs), a recently developed method that learns a self-consistent approximation of a reversed diffusion process. The CM is able to generate highly realistic global precipitation fields learned from the ERA5 reanalysis, which is employed as ground truth, in a single step. Our framework corrects the representation of extreme events as well as spatial ESM patterns especially at the small spatial scales, which are both crucial for impact assessments. Moreover, spatial biases are also corrected efficiently.  

For the specific task considered here, the CM method is \emph{up to} three orders of magnitude faster than current diffusion models based on SDEs \cite{bischoff_unpaired_2024, dhariwal_diffusion_2021, song_score-based_2021}, which need to solve a differential equation iteratively. Crucially, the CM maintains a high degree of controllability to guide the sampling in such a way that unbiased spatial patterns larger than a chosen spatial scale in ESMs are preserved. 

Similar to the SDE-based method that \cite{bischoff_unpaired_2024} apply to idealized fluid dynamics, our CM-based model only needs to be trained once on a given high-resolution observational target dataset. Since we do not condition on ESM input during training, our method can be applied to any ESM without the need for retraining. Combined with the efficient sampling of CMs, our approach is computationally fast, particularly when processing large ensembles of ESMs, without noticeable trade-offs in accuracy.

Our CM-based approach can create highly realistic fields that maintain high correlation levels with the ESMs at the spatial scales chose to be preserved. We find slightly better and at least competitive results when compared with the much more computationally expensive SDE-based method. 
The efficient single-step generation of ESM fields will be particularly relevant for processing large datasets, e.g., large ensembles as needed for uncertainty quantification, for simulations with a high temporal resolution, or long-term studies such as those in paleo-climate simulations. The ability to calibrate the ensemble sharpness and spread using different noise scales could be useful for weather prediction tasks as well. When applying our method to weather predictions, changes in the training data and splits might be of advantage to avoid distributional shifts due to the availability of satellite data.
Improving the efficiency of current deep learning models is also important from an energy consumption perspective; in this regard, our method provides a valuable contribution towards ``greener AI'' \cite{schwartz_green_2019}.

When evaluated on a future very high emission scenario, we find that our generative CM method accurately preserves the trend of the ESM, even for an extreme scenario of greenhouse gas emissions and associated global warming. This is remarkable since many machine learning-based applications to climate dynamics struggle with the out-of-sample problem imposed by our highly non-stationary climate system, when trained on historical data alone. In contrast to previous studies \cite{beucler_enforcing_2021, hess_physically_2022}, to achieve this it is not necessary to add specifically formulated physical constraints to our model. Our unconstrained CM method hence allows for a more natural generalization to unseen climate states, inherently translating non-stationary dynamics from the ESM to the downscaled high-resolution fields. However, since the global mean of the ESM is only approximately conserved with our method, adding a hard architecture constrained as in \cite{harder_hard-constrained_2023}, which fulfills a constraint up to machine precision, could further improve our results.

Grid-cell-wise autocorrelation estimates between subsequent fields suggest that CM-based stochastic downscaling method captures temporal variability of the high-resolution target data although not explicitly using temporal information. Extending the approach to temporal conditioning or predicting several time steps as done, e.g., in \cite{price_gencast_2024}, or using correlated noise as in \cite{huang_blue_2024}, might further improve the temporal dynamics.

Our downscaling method increases the resolution of the ESM simulations by a factor of four in this study. Given a higher resolution training target and more computational resources, it should, in principle, also allow for larger downscaling ratios. 

We showcase our method on uni-variate precipitation simulations because precipitation is arguably the most difficult climate variable to model. An extension to multi-variate downscaling is a natural extension of our study in future research. In principle, the convolutional CM network can be extended to include further variables as additional channels in a straightforward manner. The consistency scale will then depend on the variable (channel) and might require a separate treatment. Similarly, we believe that our method can be applied on other time scales, e.g. hourly or monthly as well, using suitably adjusted noise scales. 
Choosing the optimal noise scale using the intersection of power spectral densities as done here and in \cite{bischoff_unpaired_2024} assumes that the spectral densities are monotonically decreasing with the wavenumber. While we believe that this should hold for most climate impact variables, there might be exceptions that would require a different approach. 
We leave these explorations for future research.


\section{Methods}
\subsection{Training Data}

 As a target and ground truth dataset, we use observational precipitation data from the ERA5 reanalysis \cite{hersbach_era5_2020} provided by the European Center for Medium-Range Weather Forecasting (ECMWF).
 It covers the period from 1940 to the present, and we split the data into a training set from 1940-1990, a validation set from 1991-2003, and a test set from 2004-2018.
We bilinearly interpolate the reanalysis data to a resolution of $0.75^\circ$ and $0.9375^\circ$ in latitude and longitude direction, respectively (i.e. $240 \times 384$ grid points), which corresponds to a four times higher resolution compared to the raw ESM simulations with $3^\circ \times 3.75^\circ$ resolution (i.e. $60 \times 96)$ grid points).
 
 For the ESM precipitation fields, we use global simulations from three different ESM with varying complexity and resolution.
 The fully coupled Potsdam Earth Model (POEM)  \cite{druke_cm2mc-lpjml_2021}, which includes model components for the atmosphere, ocean, ice sheets, and dynamic vegetation, is used as the primary model for comparison of the two generative downscaling methods for past and future climates.
To demonstrate the ability of our CM-based method to correct any ESM with a coarser native resolution than the training ground truth, we further include daily precipitation simulations from the much more comprehensive and complex GFDL-ESM4 \cite{dunne_gfdl_2020}, with a native resolution of $1^\circ \times 1^\circ$. We initially upscale the GFDL-ESM4 resolution to the same grid as the POEM ESM to allow direct comparisons. We further include SpeedyWeather.jl \cite{klower_speedyweatherjl_2024}, with a native resolution of $3.75^\circ \times 3.75^\circ$, which only has a dynamic atmosphere and is, hence, less comprehensive than the fully coupled POEM ESM. Finally, we also use ERA5 data upscaled to the native POEM resolution as test data for which a paired ground truth is available. Applying our method to the latter can hence be seen as a proof of concept.

For evaluation, we use 14 years of available historical data from each of the simulations, with periods 2004-2018 for POEM, 2000-2014 for GFDL-ESM4, 1956-1970 for SpeedyWeather.jl, and 2007-2021 for ERA5.

 We apply several preprocessing steps to the simulated input data. 
 We first interpolate the input simulations onto the same high-resolution grid as the ground truth ERA5 data for downscaling purposes and model evaluation. A low-pass filter is then applied to remove small-scale artifacts created by the interpolation. Quantile delta mapping (QDM) \cite{cannon_bias_2015} with 500 quantiles is then applied in a standard way to remove distributional biases in the ESM simulation for each grid cell individually. As discussed in section \ref{sec:results_biases}, the generative downscaling only corrects biases related to a specified spatial scale. Hence, the QDM step ensures a strong reduction of single-cell biases, while the generative downscaling corrects spatial patterns that are physically consistent.
 Finally, the ESMs and ERA5 data are log-transformed, $\tilde{x} = \log(x + \epsilon) - \log(\epsilon)$ with $\epsilon = 0.0001$, as in \cite{hess_physically_2022}, followed by a normalization approximately into the range [-1, 1]. 
 
\subsection{Score-based Diffusion Models}

The underlying idea of diffusion-based generative models is to learn a \emph{reverse} diffusion process from a known prior distribution $\mathbf{x}(t=T) \sim p_T$, such as a Gaussian distribution, to a target data distribution $\mathbf{x}(t=0) \sim p_0$, where $\mathbf{x} \in \mathbb{R}^d$ and $d$ is the data dimension, e.g., the number of pixels in an image.
Score-based generative diffusion models \cite{song_generative_2019, song_score-based_2021, song_denoising_2022} generalize probabilistic denoising diffusion models \cite{ho_denoising_2020, dhariwal_diffusion_2021} to continuous-time stochastic differential equations (SDEs).

In this framework, the \emph{forward} diffusion process that incrementally perturbs the data can be described as the solution of the SDE:

\begin{equation}
    \mathrm{d}\mathbf{x} =  \mu(\mathbf{x}, t) \mathrm{d}t + g(t)\mathrm{d}\mathbf{w},
     \label{eq:forward_sde}
\end{equation}

where $\mu(\mathbf{x}, t) : \mathbb{R}^d \rightarrow  \mathbb{R}^d$  is the drift term, $\mathbf{w}$ denotes a Wiener process and $g(t) : \mathbb{R} \rightarrow  \mathbb{R}$ is the diffusion coefficient. The reverse SDE used to generate images from noise is given by \cite{song_score-based_2021} 

\begin{equation}
    \mathrm{d}\mathbf{x} = [ \mu(\mathbf{x}, t) - g(t)^2 \nabla_{\mathbf{x}} \log p_{t}(\mathbf{x}) ] \mathrm{d}\bar{t} + g(t)\mathrm{d}\mathbf{\bar{w}},
     \label{eq:reverse_sde}
\end{equation}

with $\bar{t}$ denoting a time reversal and $\nabla_{\mathbf{x}} \log p_{t}(\mathbf{x})$ being the score function of the target distribution. The score function is not analytically tractable, but one can train a score network, $s(\mathbf{x}, t; \bm{\phi}): \mathbb{R}^d \rightarrow  \mathbb{R}^d$ to approximate the score function $s(\mathbf{x}, t; \bm{\phi}) \approx \nabla_{\mathbf{x}} \log p_{t}(\mathbf{x})$, e.g., using denoising score matching \cite{song_denoising_2022} (see SI for details). For sampling, we use the Euler-Maruyama solver to integrate the reverse SDE from $t=T$ to $t=0$ in Eq.~\ref{eq:reverse_sde} with $500$ steps. 

\subsection{Consistency Models}

One major drawback of current diffusion models is that the numerical integration of the differential equation requires around 10-2000 network evaluations, depending on the solver. This makes the generation process computationally inefficient and costly compared to other generative models such as generative adversarial networks (GANs) \cite{goodfellow_generative_2014} or normalizing flows (NFs) \cite{dinh_nice_2015, papamakarios_normalizing_2021}, which can generate images in a single network evaluation. Distillation techniques can reduce the number of integration steps of diffusion models, which often represent a computational bottleneck \cite{luhman_knowledge_2021, zheng_fast_2024}.

Consistency models (CMs) can be trained from scratch without distillation and only require a single step to generate a new sample. They have been shown to outperform current distillation techniques \cite{song_consistency_2023}. 
CMs learn a \emph{consistency} function, $f (\mathbf{x}(t), t) = \mathbf{x}(t_{\mathrm{min}})$, which is self-consistent, i.e.,
\begin{equation}
f(\mathbf{x}(t), t) = f(\mathbf{x}(t'), t') \; \forall \; t, t' \in [t_{\mathrm{min}}, t_{\mathrm{max}}],
\end{equation}
where the time interval is here set to $t_{\mathrm{min}} = 0.002$ and $t_{\mathrm{max}} = 80$, following \cite{song_consistency_2023}. Further, a boundary condition $f(\mathbf{x}(t_{\mathrm{min}}), t_{\mathrm{min}}) = \mathbf{x}(t_{\mathrm{min}})$, for $t=t_{\mathrm{min}}$ is imposed. This can be implemented with the parameterization:
\begin{equation}
    f(\mathbf{x}, t; \bm{\theta}) = c_{\mathrm{skip}}(t) \mathbf{x} + c_{\mathrm{out}}(t) F(\mathbf{x}, t; \bm{\theta}),
    \label{eq:parameterization}
\end{equation}
where $F(\cdot)$ is a UNet with parameters $\bm{\theta}$. The time information is transformed using a sin-cosine positional embedding in the network. The coefficients $c_{\mathrm{skip}}(t)$ and $c_{\mathrm{out}}$ are defined, following \cite{karras_elucidating_2022, song_consistency_2023}, as
\begin{equation}
c_{\mathrm{skip}} = \frac{\sigma^2_{\mathrm{data}}}{((t-t_{\mathrm{min}})^2 + \sigma^2_{\mathrm{data}})}, \qquad  c_{\mathrm{out}}(t) = \frac{\sigma_{\mathrm{data}} t}{\sqrt{t^2 + \sigma_{\mathrm{data}^2}}}.
\label{eq:cm_coeff}
\end{equation}
The training objective is given by 
\begin{equation}
    \mathcal{L} \left( \bm{\theta},  \bm{\bar{\theta}} \right) = \mathbb{E}_{\mathbf{x},n, t_n} \left[ d \left(f(\mathbf{x} + t_{n+1}\mathbf{z}, t_{n+1}; \bm{\theta}), f \left(\mathbf{x} + t_{n}\mathbf{z}, t_{n}; \bm{ \bar{\theta}} \right) \right) \right], 
    \label{eq:loss}
\end{equation}
where $\mathbb{E}_{\mathbf{x},n, \mathbf{x}_{t_n}} \equiv \mathbb{E}_{\mathbf{x} \sim p_{\mathrm{data}},n \sim \mathcal{U}(1, N(k)-1), \mathbf{z} \sim \mathcal{N}(\mathbf{0}, \mathbf{1})}$.
The discrete time step is determined via 
\begin{equation}
t_n = \left(t_{\text{min}}^{1/\rho} + (n-1)/(N-1)(t_{\text{max}}^{1/\rho} - t_{\text{min}}^{1/\rho}) \right)^\rho,
\end{equation}
where $\rho=7$ and the discretization schedule is given by
\begin{equation}
    N(k) = \left( \sqrt{k/K ((s_1 + 1)^2 - s^2_0 ) + s^2_0} - 1 \right) + 1,
\end{equation}
where $k$ is the current training step, and $K$ is the estimated total number of training steps obtained from the PyTorch Lightning library. The initial discretization steps are set to $s_0=2$, and the maximum number of steps to $s_1=150$, following \cite{song_consistency_2023}.
With $\bm{ \bar{\theta }}$ we denote an exponential moving average (EMA) over the model parameters $\bm{\theta}$, updated with $\bm{\bar{\theta}} = \mathrm{stopgrad} [w(k) \bm{\bar{\theta}} + (1 - w(k) \bm{\theta})]$, with the decay schedule given by
\begin{equation}
    w(k) = \exp \left( \frac{s_0 \log w_0}{N(k)} \right),
\end{equation}
where $w_0=0.9$ is the initial decay rate, following \cite{song_consistency_2023}.
For the distance measure $d(\cdot, \cdot)$, we follow \cite{song_consistency_2023} and use a combination of the learned perceptual image patch similarity (LPIPS) \cite{zhang_unreasonable_2018} and $l^1$ norm:
\begin{equation}
    d(\mathbf{x},\mathbf{y}) = \mathrm{LPIPS}(\mathbf{x},\mathbf{y}) + ||\mathbf{x}-\mathbf{y}||_1.
\end{equation}
Thus, the training of the CM is self-supervised and closely related to representation learning \cite{lessig_atmorep_2023}, where a so-called "online network" $f(\cdot: \bm{\theta})$ is trained to predict the same image representation as a "target network" $f(\cdot: \bar{\bm{\theta}})$ \cite{grill_bootstrap_2020}. Importantly, the CM is thereby not trained explicitly for the downscaling tasks, which are purely performed at the inference stage.

\subsection{Network Architectures and Training}

We use a 2D UNet \cite{ronneberger_u-net_2015, song_score-based_2021} from the Diffusers library to train both the score and consistency networks from scratch, with four down- and upsampling layers. For the four layers, we use convolutions with 128, 128, 256, and 256 channels, respectively, and $3 \times 3$ kernels, SiLU activations, group normalization, and an attention layer at the architecture bottleneck. The network has, in total, around 27M trainable parameters.

We train the score network with the ADAM optimizer \cite{kingma_adam_2015} for 200 epochs, with a batch size of $1$, a learning rate of $2e^{-4}$, and an exponential moving average (EMA) over the model weights with a decay rate of 0.999 (see SI for more details).

The CM model is trained for 150 epochs following \cite{song_consistency_2023}, with the RADAM optimizer \cite{liu_variance_2021} and the same batch size, learning rate, and EMA schedule (with an initial decay rate of $\mu_0=0.9$) as the score network. We find that the loss decreases in a stable way throughout the training (Fig.~S1). The training of 150 epochs takes around six and a half days for the CM and four and a half days for the SDE on a NVIDIA V100 32GB GPU. A summary of the training hyperparameters is given in the SI in Tab.~S1.

\subsection{Scale-Consistent Downscaling}

As shown in \cite{rissanen_generative_2023, bischoff_unpaired_2024}, adding Gaussian noise with a chosen variance to an image (or fluid dynamical snapshot) results in removing spatial patterns up to a specific spatial scale associated with the amount of added noise. The trained generative model can then replace the noise with spatial patterns learned from the training data up to the chosen spatial scale.

In principle, the spatial scale can be chosen depending on the given downscaling task, e.g. related to the ESM resolution or variable. Hence, our method allows for much more flexibility after training, where the optimal spatial scale could be defined with respect to any given metric. In general, ESM fields are too smooth at small spatial scales, which presents a key problem for Earth system modelling in general and impact assessments in particular. 
More specifically, when comparing the frequency distribution of spatial precipitation fields in terms of spatial power spectral densities (PSDs), it can be seen that ESMs lack the high-frequency spatial variability, or spatial intermittency, that is a key characteristic of precipitation \cite{hess_physically_2022}. Hence, a natural choice for the spatial scale to be preserved in the ESM fields is the intersection of the PSDs from the ESM and the ground truth ERA5 \cite{bischoff_unpaired_2024} (see Fig.~\ref{fig:psds}), i.e., the scale where the ESM fields become too smooth.

For Gaussian noise, the variance as a function of time $t$ can be related to the PSD of a given wavenumber $k$ and the grid size $N$ by \cite{bischoff_unpaired_2024}

\begin{equation}
    \sigma^2(t) = N^2 \mathrm{PSD}(k).
    \label{eq:stop_time}
\end{equation}

Using Eq.~\ref{eq:stop_time}, we choose $k^*=0.0667$ (see Fig.~S2), such that it represents the wavenumber or spatial scale where the PSDs of the ESM and ERA5 precipitation fields intersect. This corresponds to  $t^*=0.468$ for the CM variance schedule and $t^*=0.355$ for the SDE bridge. 

The diffusion bridge (DB) \cite{bischoff_unpaired_2024} starts with the forward SDE in Eq.~\ref{eq:forward_sde}, initialized with a precipitation field from the POEM ESM. The forward SDE is then integrated until $t = t^*$. The reverse SDE (Eq.~\ref{eq:reverse_sde}), initialized at $t = t^*$, then denoises the field again, adding structure from the target ERA5 distribution. 

For the CM approach, at inference we apply the ``stroke guidance'' technique \cite{meng_sdedit_2022, song_consistency_2023}, where we first sample a noised ESM field $\mathbf{\tilde{x}}^{\mathrm{ESM}} \in \mathbb{R}^d$ with variance corresponding to $t^*$, 

\begin{equation}
    \mathbf{\tilde{x}}^{\mathrm{ESM}} \sim \mathcal{N}(\mathbf{x}^{\mathrm{ESM}}; \sigma^2(t^*) \mathbf{1}),
    \label{eq:stroke_guidance_noise}
\end{equation}

which is then denoised in a single step with the CM,

\begin{equation}
    \mathbf{\hat{x}} = f(\mathbf{\tilde{x}}^{\mathrm{ESM}}, t^*; \bm{\theta}),
    \label{eq:stroke_guidance_denoise}
\end{equation}

thus highly efficiently generating realistic samples $\mathbf{\hat{x}}$ that preserve unbiased spatial patterns of the ESM up to scale $k^*$.

\section*{Data Availability}
The ERA5 reanalysis data is available for download at the Copernicus Climate Change Service (C3S) 
(\url{https://cds.climate.copernicus.eu/datasets/reanalysis-era5-single-levels?tab=download}).
The simulation data from the POEM ESM is available at \url{https://doi.org/10.5281/zenodo.4683086} \cite{data-gmd-2020-436}.

\section*{Code Availability}
The Python code for processing and analyzing the data, together with the PyTorch \cite{paszke_pytorch_2019} code for training \cite{code-natmi-2024}, will be made available on GitHub: \url{https://github.com/p-hss/consistency-climate-downscaling.git}, and CodeOcean \url{https://doi.org/10.24433/CO.2150269.v1}\cite{codeocean-natmi-2024}.

\section*{Acknowledgments}
The authors would like to thank the three anonymous referees for their valuable remarks and suggestions.
The authors thank Katherine Deck and Milan Klöwer for their valuable input.
NB and PH acknowledge funding by the Volkswagen Foundation, as well as the European Regional Development Fund (ERDF), the German Federal Ministry of Education and Research, and the Land Brandenburg for supporting this project by providing resources on the high-performance computer system at the Potsdam Institute for Climate Impact Research. 
MA acknowledges funding under the Excellence Strategy of the Federal Government and the L\"ander through the TUM Innovation Network EarthCare. This is ClimTip contribution \#X; the ClimTip project has received funding from the European Union's Horizon Europe research and innovation programme under grant agreement No. 101137601.
BP acknowledges funding by the National Key R\&D Program of China (Grant NoS. 2023YFC3007700).


\section*{Author contributions}
PH and NB conceived the research and designed the study with
input from MA and BP. PH performed the model training and numerical analysis.
PH, MA, BP and NB interpreted and discussed the results. PH wrote the manuscript with input from MA, BP and NB.

\section*{Competing interests}
The authors declare no competing interests.

\printbibliography

@Misc{data-gmd-2020-436,
  author =       "Markus Dr{\"u}ke",
  title =        "{Output data for the GMD publication gmd-2020-436}",
  howpublished = "[Data set] Zenodo",
  month =        apr,
  year =         "2021",
  OPTnote =         "{\url{http://doi.org/10.5281/zenodo.4683086}}",
  url =          "http://doi.org/10.5281/zenodo.4683086",
  OPTannote =    "{Output data used in the GMD publication: CM2Mc-LPJmL v1.0: Biophysical coupling of a process-based dynamic vegetation model with managed land to a general circulation model, Dr{\"u}ke et al. (https://doi.org/10.5194/gmd-2020-436)}",
  DOI =          "10.5281/zenodo.4683086",
}

@Misc{code-natmi-2024,
  author =       "Phiipp Hess",
  title =        "{Python source code}",
  howpublished = "[Source code] Zenodo",
  year =         "2024",
  OPTnote =         "{\url{https://doi.org/10.5281/zenodo.14203092}}",
  url =          "https://doi.org/10.5281/zenodo.14203092",
  OPTannote =    "{Source code used in the publication: Fast, Scale-Adaptive, and Uncertainty-Aware Downscaling of Earth System Model Fields with Generative Machine Learning by Hess et al.",
  DOI =          "10.5281/zenodo.14203092}",
}

@Misc{codeocean-natmi-2024,
  author =       "Phiipp Hess",
  title =        "{Python source code}",
  howpublished = "[Source code] CodeOcean",
  year =         "2024",
  OPTnote =         "{\url{https://doi.org/10.24433/CO.2150269.v1}}",
  url =          "https://doi.org/10.24433/CO.2150269.v1",
  OPTannote =    "{Source code used in the publication: Fast, Scale-Adaptive, and Uncertainty-Aware Downscaling of Earth System Model Fields with Generative Machine Learning by Hess et al.",
  DOI =          "10.24433/CO.2150269.v1}",
}

@inproceedings{sauer_adversarial_2024,
	address = {Cham},
	title = {Adversarial {Diffusion} {Distillation}},
	isbn = {978-3-031-73016-0},
	doi = {10.1007/978-3-031-73016-0_6},
	language = {en},
	booktitle = {Computer {Vision} – {ECCV} 2024},
	publisher = {Springer Nature Switzerland},
	author = {Sauer, Axel and Lorenz, Dominik and Blattmann, Andreas and Rombach, Robin},
	editor = {Leonardis, Aleš and Ricci, Elisa and Roth, Stefan and Russakovsky, Olga and Sattler, Torsten and Varol, Gül},
	year = {2024},
	pages = {87--103},
}

@techreport{ecmwf_ifs_2016,
	title = {{IFS} {Documentation} {CY41R2} - {Part} {IV}: {Physical} {Processes}},
	shorttitle = {{IFS} {Documentation} {CY41R2} - {Part} {IV}},
	url = {https://www.ecmwf.int/en/elibrary/79697-ifs-documentation-cy41r2-part-iv-physical-processes},
	language = {eng},
	number = {2},
	urldate = {2023-08-29},
	institution = {ECMWF},
	author = {ECMWF},
	year = {2016},
}

@article{von_bloh_implementing_2018,
	title = {Implementing the nitrogen cycle into the dynamic global vegetation, hydrology, and crop growth model {LPJmL} (version 5.0)},
	volume = {11},
	issn = {1991-959X},
	url = {https://gmd.copernicus.org/articles/11/2789/2018/},
	doi = {10.5194/gmd-11-2789-2018},
	language = {English},
	number = {7},
	urldate = {2024-11-18},
	journal = {Geoscientific Model Development},
	author = {von Bloh, Werner and Schaphoff, Sibyll and Müller, Christoph and Rolinski, Susanne and Waha, Katharina and Zaehle, Sönke},
	month = jul,
	year = {2018},
	note = {Publisher: Copernicus GmbH},
	pages = {2789--2812},
}

@article{galbraith_climate_2011,
	title = {Climate {Variability} and {Radiocarbon} in the {CM2Mc} {Earth} {System} {Model}},
	volume = {24},
	issn = {0894-8755, 1520-0442},
	url = {https://journals.ametsoc.org/view/journals/clim/24/16/2011jcli3919.1.xml},
	doi = {10.1175/2011JCLI3919.1},
	language = {EN},
	number = {16},
	urldate = {2024-11-18},
	journal = {Journal of Climate},
	author = {Galbraith, Eric D. and Kwon, Eun Young and Gnanadesikan, Anand and Rodgers, Keith B. and Griffies, Stephen M. and Bianchi, Daniele and Sarmiento, Jorge L. and Dunne, John P. and Simeon, Jennifer and Slater, Richard D. and Wittenberg, Andrew T. and Held, Isaac M.},
	month = aug,
	year = {2011},
	note = {Publisher: American Meteorological Society
Section: Journal of Climate},
	pages = {4230--4254},
}

@article{harder_hard-constrained_2023,
	title = {Hard-{Constrained} {Deep} {Learning} for {Climate} {Downscaling}},
	volume = {24},
	issn = {1533-7928},
	url = {http://jmlr.org/papers/v24/23-0158.html},
	number = {365},
	urldate = {2024-11-14},
	journal = {Journal of Machine Learning Research},
	author = {Harder, Paula and Hernandez-Garcia, Alex and Ramesh, Venkatesh and Yang, Qidong and Sattegeri, Prasanna and Szwarcman, Daniela and Watson, Campbell and Rolnick, David},
	year = {2023},
	pages = {1--40},
}

@article{bischoff_unpaired_2024,
	title = {Unpaired {Downscaling} of {Fluid} {Flows} with {Diffusion} {Bridges}},
	volume = {3},
	issn = {2769-7525},
	url = {https://journals.ametsoc.org/view/journals/aies/3/2/AIES-D-23-0039.1.xml},
	doi = {10.1175/AIES-D-23-0039.1},
	language = {EN},
	number = {2},
	urldate = {2024-11-14},
	journal = {Artificial Intelligence for the Earth Systems},
	author = {Bischoff, Tobias and Deck, Katherine},
	month = may,
	year = {2024},
	note = {Publisher: American Meteorological Society
Section: Artificial Intelligence for the Earth Systems},
}

@article{clark_text--image_2023,
	title = {Text-to-{Image} {Diffusion} {Models} are {Zero} {Shot} {Classifiers}},
	volume = {36},
	url = {https://proceedings.neurips.cc/paper_files/paper/2023/hash/b87bdcf963cad3d0b265fcb78ae7d11e-Abstract-Conference.html},
	language = {en},
	urldate = {2024-07-16},
	journal = {Advances in Neural Information Processing Systems},
	author = {Clark, Kevin and Jaini, Priyank},
	month = dec,
	year = {2023},
	pages = {58921--58937},
}

@inproceedings{grill_bootstrap_2020,
	title = {Bootstrap {Your} {Own} {Latent} - {A} {New} {Approach} to {Self}-{Supervised} {Learning}},
	volume = {33},
	url = {https://proceedings.neurips.cc/paper/2020/hash/f3ada80d5c4ee70142b17b8192b2958e-Abstract.html},
	urldate = {2024-07-16},
	booktitle = {Advances in {Neural} {Information} {Processing} {Systems}},
	publisher = {Curran Associates, Inc.},
	author = {Grill, Jean-Bastien and Strub, Florian and Altché, Florent and Tallec, Corentin and Richemond, Pierre and Buchatskaya, Elena and Doersch, Carl and Avila Pires, Bernardo and Guo, Zhaohan and Gheshlaghi Azar, Mohammad and Piot, Bilal and kavukcuoglu, koray and Munos, Remi and Valko, Michal},
	year = {2020},
	pages = {21271--21284},
}

@misc{lessig_atmorep_2023,
	title = {{AtmoRep}: {A} stochastic model of atmosphere dynamics using large scale representation learning},
	shorttitle = {{AtmoRep}},
	url = {http://arxiv.org/abs/2308.13280},
	doi = {10.48550/arXiv.2308.13280},
	urldate = {2024-07-16},
	publisher = {arXiv},
	author = {Lessig, Christian and Luise, Ilaria and Gong, Bing and Langguth, Michael and Stadtler, Scarlet and Schultz, Martin},
	month = sep,
	year = {2023},
	note = {arXiv:2308.13280 [physics]},
}

@inproceedings{song_consistency_2023,
	title = {Consistency {Models}},
	url = {https://proceedings.mlr.press/v202/song23a.html},
	language = {en},
	urldate = {2024-07-15},
	booktitle = {Proceedings of the 40th {International} {Conference} on {Machine} {Learning}},
	publisher = {PMLR},
	author = {Song, Yang and Dhariwal, Prafulla and Chen, Mark and Sutskever, Ilya},
	month = jul,
	year = {2023},
	note = {ISSN: 2640-3498},
	pages = {32211--32252},
}

@article{klower_speedyweatherjl_2024,
	title = {{SpeedyWeather}.jl: {Reinventing} atmospheric generalcirculation models towards interactivity and extensibility},
	volume = {9},
	copyright = {http://creativecommons.org/licenses/by/4.0/},
	issn = {2475-9066},
	shorttitle = {{SpeedyWeather}.jl},
	url = {https://joss.theoj.org/papers/10.21105/joss.06323},
	doi = {10.21105/joss.06323},
	language = {en},
	number = {98},
	urldate = {2024-07-11},
	journal = {Journal of Open Source Software},
	author = {Klöwer, Milan and Gelbrecht, Maximilian and Hotta, Daisuke and Willmert, Justin and Silvestri, Simone and Wagner, Gregory L and White, Alistair and Hatfield, Sam and Kimpson, Tom and Constantinou, Navid C and Hill, Chris},
	month = jun,
	year = {2024},
	pages = {6323},
}

@article{gneiting_probabilistic_2007,
	title = {Probabilistic forecasts, calibration and sharpness},
	volume = {69},
	issn = {1467-9868},
	url = {https://onlinelibrary.wiley.com/doi/abs/10.1111/j.1467-9868.2007.00587.x},
	doi = {10.1111/j.1467-9868.2007.00587.x},
	language = {en},
	number = {2},
	urldate = {2022-11-03},
	journal = {Journal of the Royal Statistical Society: Series B (Statistical Methodology)},
	author = {Gneiting, Tilmann and Balabdaoui, Fadoua and Raftery, Adrian E.},
	year = {2007},
	note = {\_eprint: https://onlinelibrary.wiley.com/doi/pdf/10.1111/j.1467-9868.2007.00587.x},
	pages = {243--268},
}

@article{eyring_overview_2016,
	title = {Overview of the {Coupled} {Model} {Intercomparison} {Project} {Phase} 6 ({CMIP6}) experimental design and organization},
	volume = {9},
	issn = {1991-959X},
	url = {https://gmd.copernicus.org/articles/9/1937/2016/},
	doi = {10.5194/gmd-9-1937-2016},
	language = {English},
	number = {5},
	urldate = {2024-06-27},
	journal = {Geoscientific Model Development},
	author = {Eyring, Veronika and Bony, Sandrine and Meehl, Gerald A. and Senior, Catherine A. and Stevens, Bjorn and Stouffer, Ronald J. and Taylor, Karl E.},
	month = may,
	year = {2016},
	note = {Publisher: Copernicus GmbH},
	pages = {1937--1958},
}

@misc{price_gencast_2024,
	title = {{GenCast}: {Diffusion}-based ensemble forecasting for medium-range weather},
	shorttitle = {{GenCast}},
	url = {http://arxiv.org/abs/2312.15796},
	doi = {10.48550/arXiv.2312.15796},
	urldate = {2024-05-07},
	publisher = {arXiv},
	author = {Price, Ilan and Sanchez-Gonzalez, Alvaro and Alet, Ferran and Andersson, Tom R. and El-Kadi, Andrew and Masters, Dominic and Ewalds, Timo and Stott, Jacklynn and Mohamed, Shakir and Battaglia, Peter and Lam, Remi and Willson, Matthew},
	month = may,
	year = {2024},
	note = {arXiv:2312.15796 [physics]},
}

@misc{zheng_fast_2024,
	title = {Fast {Training} of {Diffusion} {Models} with {Masked} {Transformers}},
	url = {http://arxiv.org/abs/2306.09305},
	language = {en},
	urldate = {2024-06-26},
	publisher = {arXiv},
	author = {Zheng, Hongkai and Nie, Weili and Vahdat, Arash and Anandkumar, Anima},
	month = mar,
	year = {2024},
	note = {arXiv:2306.09305 [cs]},
}

@misc{huang_blue_2024,
	title = {Blue noise for diffusion models},
	url = {http://arxiv.org/abs/2402.04930},
	language = {en},
	urldate = {2024-06-17},
	publisher = {arXiv},
	author = {Huang, Xingchang and Salaün, Corentin and Vasconcelos, Cristina and Theobalt, Christian and Öztireli, Cengiz and Singh, Gurprit},
	month = may,
	year = {2024},
	note = {arXiv:2402.04930 [cs]},
}

@article{hess_deep_2023,
	title = {Deep {Learning} for {Bias}-{Correcting} {CMIP6}-{Class} {Earth} {System} {Models}},
	volume = {11},
	copyright = {© 2023 The Authors.},
	issn = {2328-4277},
	url = {https://onlinelibrary.wiley.com/doi/abs/10.1029/2023EF004002},
	doi = {10.1029/2023EF004002},
	language = {en},
	number = {10},
	urldate = {2023-12-04},
	journal = {Earth's Future},
	author = {Hess, Philipp and Lange, Stefan and Schötz, Christof and Boers, Niklas},
	year = {2023},
	pages = {e2023EF004002},
}

@misc{esser_scaling_2024,
	title = {Scaling {Rectified} {Flow} {Transformers} for {High}-{Resolution} {Image} {Synthesis}},
	url = {http://arxiv.org/abs/2403.03206},
	language = {en},
	urldate = {2024-04-09},
	publisher = {arXiv},
	author = {Esser, Patrick and Kulal, Sumith and Blattmann, Andreas and Entezari, Rahim and Müller, Jonas and Saini, Harry and Levi, Yam and Lorenz, Dominik and Sauer, Axel and Boesel, Frederic and Podell, Dustin and Dockhorn, Tim and English, Zion and Lacey, Kyle and Goodwin, Alex and Marek, Yannik and Rombach, Robin},
	month = mar,
	year = {2024},
	note = {arXiv:2403.03206 [cs]},
}

@misc{liu_variance_2021,
	title = {On the {Variance} of the {Adaptive} {Learning} {Rate} and {Beyond}},
	url = {http://arxiv.org/abs/1908.03265},
	doi = {10.48550/arXiv.1908.03265},
	urldate = {2024-01-30},
	publisher = {arXiv},
	author = {Liu, Liyuan and Jiang, Haoming and He, Pengcheng and Chen, Weizhu and Liu, Xiaodong and Gao, Jianfeng and Han, Jiawei},
	month = oct,
	year = {2021},
	note = {arXiv:1908.03265 [cs, stat]},
}

@misc{dinh_nice_2015,
	title = {{NICE}: {Non}-linear {Independent} {Components} {Estimation}},
	shorttitle = {{NICE}},
	url = {http://arxiv.org/abs/1410.8516},
	doi = {10.48550/arXiv.1410.8516},
	urldate = {2024-01-30},
	publisher = {arXiv},
	author = {Dinh, Laurent and Krueger, David and Bengio, Yoshua},
	month = apr,
	year = {2015},
	note = {arXiv:1410.8516 [cs]},
}

@misc{luhman_knowledge_2021,
	title = {Knowledge {Distillation} in {Iterative} {Generative} {Models} for {Improved} {Sampling} {Speed}},
	url = {http://arxiv.org/abs/2101.02388},
	doi = {10.48550/arXiv.2101.02388},
	urldate = {2024-01-30},
	publisher = {arXiv},
	author = {Luhman, Eric and Luhman, Troy},
	month = jan,
	year = {2021},
	note = {arXiv:2101.02388 [cs]},
}

@article{papamakarios_normalizing_2021,
	title = {Normalizing flows for probabilistic modeling and inference},
	volume = {22},
	issn = {1532-4435},
	number = {1},
	journal = {The Journal of Machine Learning Research},
	author = {Papamakarios, George and Nalisnick, Eric and Rezende, Danilo Jimenez and Mohamed, Shakir and Lakshminarayanan, Balaji},
	month = jan,
	year = {2021},
	pages = {57:2617--57:2680},
}

@misc{schwartz_green_2019,
	title = {Green {AI}},
	url = {http://arxiv.org/abs/1907.10597},
	language = {en},
	urldate = {2024-01-26},
	publisher = {arXiv},
	author = {Schwartz, Roy and Dodge, Jesse and Smith, Noah A. and Etzioni, Oren},
	month = aug,
	year = {2019},
	note = {arXiv:1907.10597 [cs, stat]},
}

@misc{rissanen_generative_2023,
	title = {Generative {Modelling} {With} {Inverse} {Heat} {Dissipation}},
	url = {http://arxiv.org/abs/2206.13397},
	language = {en},
	urldate = {2023-06-01},
	publisher = {arXiv},
	author = {Rissanen, Severi and Heinonen, Markus and Solin, Arno},
	month = apr,
	year = {2023},
	note = {arXiv:2206.13397 [cs, stat]},
}

@article{vincent_connection_2011,
	title = {A {Connection} {Between} {Score} {Matching} and {Denoising} {Autoencoders}},
	volume = {23},
	issn = {0899-7667},
	url = {https://ieeexplore.ieee.org/abstract/document/6795935},
	doi = {10.1162/NECO_a_00142},
	number = {7},
	urldate = {2023-12-01},
	journal = {Neural Computation},
	author = {Vincent, Pascal},
	month = jul,
	year = {2011},
	note = {Conference Name: Neural Computation},
	pages = {1661--1674},
}

@article{anderson_reverse-time_1982,
	title = {Reverse-time diffusion equation models},
	volume = {12},
	issn = {0304-4149},
	url = {https://www.sciencedirect.com/science/article/pii/0304414982900515},
	doi = {10.1016/0304-4149(82)90051-5},
	number = {3},
	urldate = {2023-12-01},
	journal = {Stochastic Processes and their Applications},
	author = {Anderson, Brian D. O.},
	month = may,
	year = {1982},
	pages = {313--326},
}

@article{hersbach_era5_2020,
	title = {The {ERA5} global reanalysis},
	volume = {146},
	copyright = {© 2020 The Authors. Quarterly Journal of the Royal Meteorological Society published by John Wiley \& Sons Ltd on behalf of the Royal Meteorological Society.},
	issn = {1477-870X},
	url = {https://onlinelibrary.wiley.com/doi/abs/10.1002/qj.3803},
	doi = {10.1002/qj.3803},
	language = {en},
	number = {730},
	urldate = {2023-08-24},
	journal = {Quarterly Journal of the Royal Meteorological Society},
	author = {Hersbach, Hans and Bell, Bill and Berrisford, Paul and Hirahara, Shoji and Horányi, András and Muñoz-Sabater, Joaquín and Nicolas, Julien and Peubey, Carole and Radu, Raluca and Schepers, Dinand and Simmons, Adrian and Soci, Cornel and Abdalla, Saleh and Abellan, Xavier and Balsamo, Gianpaolo and Bechtold, Peter and Biavati, Gionata and Bidlot, Jean and Bonavita, Massimo and De Chiara, Giovanna and Dahlgren, Per and Dee, Dick and Diamantakis, Michail and Dragani, Rossana and Flemming, Johannes and Forbes, Richard and Fuentes, Manuel and Geer, Alan and Haimberger, Leo and Healy, Sean and Hogan, Robin J. and Hólm, Elías and Janisková, Marta and Keeley, Sarah and Laloyaux, Patrick and Lopez, Philippe and Lupu, Cristina and Radnoti, Gabor and de Rosnay, Patricia and Rozum, Iryna and Vamborg, Freja and Villaume, Sebastien and Thépaut, Jean-Noël},
	year = {2020},
	note = {\_eprint: https://onlinelibrary.wiley.com/doi/pdf/10.1002/qj.3803},
	pages = {1999--2049},
}

@inproceedings{song_generative_2019,
	title = {Generative {Modeling} by {Estimating} {Gradients} of the {Data} {Distribution}},
	volume = {32},
	url = {https://proceedings.neurips.cc/paper_files/paper/2019/hash/3001ef257407d5a371a96dcd947c7d93-Abstract.html},
	urldate = {2023-12-01},
	booktitle = {Advances in {Neural} {Information} {Processing} {Systems}},
	publisher = {Curran Associates, Inc.},
	author = {Song, Yang and Ermon, Stefano},
	year = {2019},
}

@article{pan_learning_2021,
	title = {Learning to {Correct} {Climate} {Projection} {Biases}},
	volume = {13},
	copyright = {© 2021 The Authors. Journal of Advances in Modeling Earth Systems published by Wiley Periodicals LLC on behalf of American Geophysical Union.},
	issn = {1942-2466},
	url = {https://onlinelibrary.wiley.com/doi/abs/10.1029/2021MS002509},
	doi = {10.1029/2021MS002509},
	language = {en},
	number = {10},
	urldate = {2023-12-01},
	journal = {Journal of Advances in Modeling Earth Systems},
	author = {Pan, Baoxiang and Anderson, Gemma J. and Goncalves, André and Lucas, Donald D. and Bonfils, Céline J. W. and Lee, Jiwoo and Tian, Yang and Ma, Hsi-Yen},
	year = {2021},
	note = {\_eprint: https://onlinelibrary.wiley.com/doi/pdf/10.1029/2021MS002509},
	pages = {e2021MS002509},
}

@misc{zhang_unreasonable_2018,
	title = {The {Unreasonable} {Effectiveness} of {Deep} {Features} as a {Perceptual} {Metric}},
	url = {http://arxiv.org/abs/1801.03924},
	language = {en},
	urldate = {2023-07-24},
	publisher = {arXiv},
	author = {Zhang, Richard and Isola, Phillip and Efros, Alexei A. and Shechtman, Eli and Wang, Oliver},
	month = apr,
	year = {2018},
	note = {arXiv:1801.03924 [cs]},
}

@misc{wan_debias_2023,
	title = {Debias {Coarsely}, {Sample} {Conditionally}: {Statistical} {Downscaling} through {Optimal} {Transport} and {Probabilistic} {Diffusion} {Models}},
	shorttitle = {Debias {Coarsely}, {Sample} {Conditionally}},
	url = {http://arxiv.org/abs/2305.15618},
	language = {en},
	urldate = {2023-06-22},
	publisher = {arXiv},
	author = {Wan, Zhong Yi and Baptista, Ricardo and Chen, Yi-fan and Anderson, John and Boral, Anudhyan and Sha, Fei and Zepeda-Núñez, Leonardo},
	month = may,
	year = {2023},
	note = {arXiv:2305.15618 [physics]},
}

@misc{karras_elucidating_2022,
	title = {Elucidating the {Design} {Space} of {Diffusion}-{Based} {Generative} {Models}},
	url = {http://arxiv.org/abs/2206.00364},
	language = {en},
	urldate = {2023-05-09},
	publisher = {arXiv},
	author = {Karras, Tero and Aittala, Miika and Aila, Timo and Laine, Samuli},
	month = oct,
	year = {2022},
	note = {arXiv:2206.00364 [cs, stat]},
}

@misc{meng_sdedit_2022,
	title = {{SDEdit}: {Guided} {Image} {Synthesis} and {Editing} with {Stochastic} {Differential} {Equations}},
	shorttitle = {{SDEdit}},
	url = {http://arxiv.org/abs/2108.01073},
	language = {en},
	urldate = {2023-04-19},
	publisher = {arXiv},
	author = {Meng, Chenlin and He, Yutong and Song, Yang and Song, Jiaming and Wu, Jiajun and Zhu, Jun-Yan and Ermon, Stefano},
	month = jan,
	year = {2022},
	note = {arXiv:2108.01073 [cs]},
}

@misc{song_score-based_2021,
	title = {Score-{Based} {Generative} {Modeling} through {Stochastic} {Differential} {Equations}},
	url = {http://arxiv.org/abs/2011.13456},
	language = {en},
	urldate = {2023-04-18},
	publisher = {arXiv},
	author = {Song, Yang and Sohl-Dickstein, Jascha and Kingma, Diederik P. and Kumar, Abhishek and Ermon, Stefano and Poole, Ben},
	month = feb,
	year = {2021},
	note = {arXiv:2011.13456 [cs, stat]},
}

@misc{song_denoising_2022,
	title = {Denoising {Diffusion} {Implicit} {Models}},
	url = {http://arxiv.org/abs/2010.02502},
	language = {en},
	urldate = {2023-03-24},
	publisher = {arXiv},
	author = {Song, Jiaming and Meng, Chenlin and Ermon, Stefano},
	month = oct,
	year = {2022},
	note = {arXiv:2010.02502 [cs]},
}

@misc{dhariwal_diffusion_2021,
	title = {Diffusion {Models} {Beat} {GANs} on {Image} {Synthesis}},
	url = {http://arxiv.org/abs/2105.05233},
	language = {en},
	urldate = {2023-03-24},
	publisher = {arXiv},
	author = {Dhariwal, Prafulla and Nichol, Alex},
	month = jun,
	year = {2021},
	note = {arXiv:2105.05233 [cs, stat]},
}

@misc{ho_denoising_2020,
	title = {Denoising {Diffusion} {Probabilistic} {Models}},
	url = {http://arxiv.org/abs/2006.11239},
	language = {en},
	urldate = {2023-03-22},
	publisher = {arXiv},
	author = {Ho, Jonathan and Jain, Ajay and Abbeel, Pieter},
	month = dec,
	year = {2020},
	note = {arXiv:2006.11239 [cs, stat]},
}

@misc{groenke_climalign_2020,
	title = {{ClimAlign}: {Unsupervised} statistical downscaling of climate variables via normalizing flows},
	shorttitle = {{ClimAlign}},
	url = {http://arxiv.org/abs/2008.04679},
	language = {en},
	urldate = {2023-02-17},
	publisher = {arXiv},
	author = {Groenke, Brian and Madaus, Luke and Monteleoni, Claire},
	month = aug,
	year = {2020},
	note = {arXiv:2008.04679 [cs, stat]},
}

@article{druke_cm2mc-lpjml_2021,
	title = {{CM2Mc}-{LPJmL} v1.0: biophysical coupling of a process-based dynamic vegetation model with managed land to a general circulation model},
	volume = {14},
	issn = {1991-959X},
	shorttitle = {{CM2Mc}-{LPJmL} v1.0},
	url = {https://gmd.copernicus.org/articles/14/4117/2021/},
	doi = {10.5194/gmd-14-4117-2021},
	language = {English},
	number = {6},
	urldate = {2023-01-06},
	journal = {Geoscientific Model Development},
	author = {Drüke, Markus and von Bloh, Werner and Petri, Stefan and Sakschewski, Boris and Schaphoff, Sibyll and Forkel, Matthias and Huiskamp, Willem and Feulner, Georg and Thonicke, Kirsten},
	month = jul,
	year = {2021},
	note = {Publisher: Copernicus GmbH},
	pages = {4117--4141},
}

@article{harris_generative_2022,
	title = {A {Generative} {Deep} {Learning} {Approach} to {Stochastic} {Downscaling} of {Precipitation} {Forecasts}},
	volume = {14},
	issn = {1942-2466, 1942-2466},
	url = {http://arxiv.org/abs/2204.02028},
	doi = {10.1029/2022MS003120},
	number = {10},
	urldate = {2023-01-06},
	journal = {Journal of Advances in Modeling Earth Systems},
	author = {Harris, Lucy and McRae, Andrew T. T. and Chantry, Matthew and Dueben, Peter D. and Palmer, Tim N.},
	month = oct,
	year = {2022},
	note = {arXiv:2204.02028 [physics, stat]},
}

@article{dunne_gfdl_2020,
	title = {The {GFDL} {Earth} {System} {Model} {Version} 4.1 ({GFDL}-{ESM} 4.1): {Overall} {Coupled} {Model} {Description} and {Simulation} {Characteristics}},
	volume = {12},
	issn = {1942-2466},
	shorttitle = {The {GFDL} {Earth} {System} {Model} {Version} 4.1 ({GFDL}-{ESM} 4.1)},
	url = {https://onlinelibrary.wiley.com/doi/abs/10.1029/2019MS002015},
	doi = {10.1029/2019MS002015},
	language = {en},
	number = {11},
	urldate = {2022-12-30},
	journal = {Journal of Advances in Modeling Earth Systems},
	author = {Dunne, J. P. and Horowitz, L. W. and Adcroft, A. J. and Ginoux, P. and Held, I. M. and John, J. G. and Krasting, J. P. and Malyshev, S. and Naik, V. and Paulot, F. and Shevliakova, E. and Stock, C. A. and Zadeh, N. and Balaji, V. and Blanton, C. and Dunne, K. A. and Dupuis, C. and Durachta, J. and Dussin, R. and Gauthier, P. P. G. and Griffies, S. M. and Guo, H. and Hallberg, R. W. and Harrison, M. and He, J. and Hurlin, W. and McHugh, C. and Menzel, R. and Milly, P. C. D. and Nikonov, S. and Paynter, D. J. and Ploshay, J. and Radhakrishnan, A. and Rand, K. and Reichl, B. G. and Robinson, T. and Schwarzkopf, D. M. and Sentman, L. T. and Underwood, S. and Vahlenkamp, H. and Winton, M. and Wittenberg, A. T. and Wyman, B. and Zeng, Y. and Zhao, M.},
	year = {2020},
	note = {\_eprint: https://onlinelibrary.wiley.com/doi/pdf/10.1029/2019MS002015},
	pages = {e2019MS002015},
}

@article{kotz_effect_2022,
	title = {The effect of rainfall changes on economic production},
	volume = {601},
	copyright = {2022 The Author(s), under exclusive licence to Springer Nature Limited},
	issn = {1476-4687},
	url = {https://www.nature.com/articles/s41586-021-04283-8$},
	doi = {10.1038/s41586-021-04283-8},
	language = {en},
	number = {7892},
	urldate = {2022-12-02},
	journal = {Nature},
	author = {Kotz, Maximilian and Levermann, Anders and Wenz, Leonie},
	month = jan,
	year = {2022},
	note = {Number: 7892
Publisher: Nature Publishing Group},
	pages = {223--227},
}

@inproceedings{paszke_pytorch_2019,
	title = {{PyTorch}: {An} {Imperative} {Style}, {High}-{Performance} {Deep} {Learning} {Library}},
	volume = {32},
	shorttitle = {{PyTorch}},
	url = {https://proceedings.neurips.cc/paper/2019/hash/bdbca288fee7f92f2bfa9f7012727740-Abstract.html},
	urldate = {2022-12-02},
	booktitle = {Advances in {Neural} {Information} {Processing} {Systems}},
	publisher = {Curran Associates, Inc.},
	author = {Paszke, Adam and Gross, Sam and Massa, Francisco and Lerer, Adam and Bradbury, James and Chanan, Gregory and Killeen, Trevor and Lin, Zeming and Gimelshein, Natalia and Antiga, Luca and Desmaison, Alban and Kopf, Andreas and Yang, Edward and DeVito, Zachary and Raison, Martin and Tejani, Alykhan and Chilamkurthy, Sasank and Steiner, Benoit and Fang, Lu and Bai, Junjie and Chintala, Soumith},
	year = {2019},
}

@misc{arjovsky_towards_2017,
	title = {Towards {Principled} {Methods} for {Training} {Generative} {Adversarial} {Networks}},
	url = {http://arxiv.org/abs/1701.04862},
	doi = {10.48550/arXiv.1701.04862},
	urldate = {2022-11-30},
	publisher = {arXiv},
	author = {Arjovsky, Martin and Bottou, Léon},
	month = jan,
	year = {2017},
	note = {arXiv:1701.04862 [cs, stat]},
}

@inproceedings{goodfellow_generative_2014,
	title = {Generative {Adversarial} {Nets}},
	volume = {27},
	url = {https://proceedings.neurips.cc/paper/2014/file/5ca3e9b122f61f8f06494c97b1afccf3-Paper.pdf},
	booktitle = {Advances in {Neural} {Information} {Processing} {Systems}},
	publisher = {Curran Associates, Inc.},
	author = {Goodfellow, Ian and Pouget-Abadie, Jean and Mirza, Mehdi and Xu, Bing and Warde-Farley, David and Ozair, Sherjil and Courville, Aaron and Bengio, Yoshua},
	editor = {Ghahramani, Z. and Welling, M. and Cortes, C. and Lawrence, N. and Weinberger, K. Q.},
	year = {2014},
}

@article{schneider_climate_2017,
	title = {Climate goals and computing the future of clouds},
	volume = {7},
	copyright = {2017 Nature Publishing Group, a division of Macmillan Publishers Limited. All Rights Reserved.},
	issn = {1758-6798},
	url = {https://www.nature.com/articles/nclimate3190},
	doi = {10.1038/nclimate3190},
	language = {en},
	number = {1},
	urldate = {2022-11-24},
	journal = {Nature Climate Change},
	author = {Schneider, Tapio and Teixeira, João and Bretherton, Christopher S. and Brient, Florent and Pressel, Kyle G. and Schär, Christoph and Siebesma, A. Pier},
	month = jan,
	year = {2017},
	note = {Number: 1
Publisher: Nature Publishing Group},
	pages = {3--5},
}

@article{schaphoff_lpjml4_2018,
	title = {{LPJmL4} – a dynamic global vegetation model with managed land – {Part} 1: {Model} description},
	volume = {11},
	issn = {1991-959X},
	shorttitle = {{LPJmL4} – a dynamic global vegetation model with managed land – {Part} 1},
	url = {https://gmd.copernicus.org/articles/11/1343/2018/},
	doi = {10.5194/gmd-11-1343-2018},
	language = {English},
	number = {4},
	urldate = {2022-11-07},
	journal = {Geoscientific Model Development},
	author = {Schaphoff, Sibyll and von Bloh, Werner and Rammig, Anja and Thonicke, Kirsten and Biemans, Hester and Forkel, Matthias and Gerten, Dieter and Heinke, Jens and Jägermeyr, Jonas and Knauer, Jürgen and Langerwisch, Fanny and Lucht, Wolfgang and Müller, Christoph and Rolinski, Susanne and Waha, Katharina},
	month = apr,
	year = {2018},
	note = {Publisher: Copernicus GmbH},
	pages = {1343--1375},
}

@article{traxl_role_2021,
	title = {The role of cyclonic activity in tropical temperature-rainfall scaling},
	volume = {12},
	copyright = {2021 The Author(s)},
	issn = {2041-1723},
	url = {https://www.nature.com/articles/s41467-021-27111-z},
	doi = {10.1038/s41467-021-27111-z},
	language = {en},
	number = {1},
	urldate = {2022-11-04},
	journal = {Nature Communications},
	author = {Traxl, Dominik and Boers, Niklas and Rheinwalt, Aljoscha and Bookhagen, Bodo},
	month = nov,
	year = {2021},
	note = {Number: 1
Publisher: Nature Publishing Group},
	pages = {6732},
}

@article{cannon_bias_2015,
	title = {Bias {Correction} of {GCM} {Precipitation} by {Quantile} {Mapping}: {How} {Well} {Do} {Methods} {Preserve} {Changes} in {Quantiles} and {Extremes}?},
	volume = {28},
	issn = {0894-8755, 1520-0442},
	shorttitle = {Bias {Correction} of {GCM} {Precipitation} by {Quantile} {Mapping}},
	url = {https://journals.ametsoc.org/view/journals/clim/28/17/jcli-d-14-00754.1.xml},
	doi = {10.1175/JCLI-D-14-00754.1},
	language = {EN},
	number = {17},
	urldate = {2022-10-28},
	journal = {Journal of Climate},
	author = {Cannon, Alex J. and Sobie, Stephen R. and Murdock, Trevor Q.},
	month = sep,
	year = {2015},
	note = {Publisher: American Meteorological Society
Section: Journal of Climate},
	pages = {6938--6959},
}

@article{hess_physically_2022,
	title = {Physically constrained generative adversarial networks for improving precipitation fields from {Earth} system models},
	volume = {4},
	copyright = {2022 The Author(s), under exclusive licence to Springer Nature Limited},
	issn = {2522-5839},
	url = {https://www.nature.com/articles/s42256-022-00540-1},
	doi = {10.1038/s42256-022-00540-1},
	language = {en},
	number = {10},
	urldate = {2022-10-27},
	journal = {Nature Machine Intelligence},
	author = {Hess, Philipp and Drüke, Markus and Petri, Stefan and Strnad, Felix M. and Boers, Niklas},
	month = oct,
	year = {2022},
	note = {Number: 10
Publisher: Nature Publishing Group},
	pages = {828--839},
}

@article{kingma_adam_2015,
	title = {Adam: {A} method for stochastic optimization},
	journal = {3rd International Conference on Learning Representations, ICLR 2015 - Conference Track Proceedings},
	author = {Kingma, Diederik P. and Ba, Jimmy Lei},
	year = {2015},
	note = {arXiv: 1412.6980},
	pages = {1--15},
}

@article{ravuri_skilful_2021,
	title = {Skilful precipitation nowcasting using deep generative models of radar},
	volume = {597},
	issn = {14764687},
	url = {http://arxiv.org/abs/2104.00954},
	doi = {10.1038/s41586-021-03854-z},
	number = {7878},
	journal = {Nature},
	author = {Ravuri, Suman and Lenc, Karel and Willson, Matthew and Kangin, Dmitry and Lam, Remi and Mirowski, Piotr and Fitzsimons, Megan and Athanassiadou, Maria and Kashem, Sheleem and Madge, Sam and Prudden, Rachel and Mandhane, Amol and Clark, Aidan and Brock, Andrew and Simonyan, Karen and Hadsell, Raia and Robinson, Niall and Clancy, Ellen and Arribas, Alberto and Mohamed, Shakir},
	year = {2021},
	pmid = {34588668},
	note = {arXiv: 2104.00954
Publisher: Springer US},
	pages = {672--677},
}

@inproceedings{ronneberger_u-net_2015,
	title = {U-net: {Convolutional} networks for biomedical image segmentation},
	volume = {9351},
	isbn = {978-3-319-24573-7},
	url = {http://lmb.informatik.uni-freiburg.de/},
	doi = {10.1007/978-3-319-24574-4_28},
	urldate = {2020-04-26},
	booktitle = {Lecture {Notes} in {Computer} {Science} (including subseries {Lecture} {Notes} in {Artificial} {Intelligence} and {Lecture} {Notes} in {Bioinformatics})},
	publisher = {Springer Verlag},
	author = {Ronneberger, Olaf and Fischer, Philipp and Brox, Thomas},
	month = may,
	year = {2015},
	note = {arXiv: 1505.04597
ISSN: 16113349},
	pages = {234--241},
}

@article{hersbach_decomposition_2000,
	title = {Decomposition of the continuous ranked probability score for ensemble prediction systems},
	volume = {15},
	issn = {08828156},
	url = {https://journals.ametsoc.org/view/journals/wefo/15/5/1520-0434_2000_015_0559_dotcrp_2_0_co_2.xml},
	doi = {10.1175/1520-0434(2000)015<0559:DOTCRP>2.0.CO;2},
	number = {5},
	urldate = {2021-03-09},
	journal = {Weather and Forecasting},
	author = {Hersbach, H.},
	month = oct,
	year = {2000},
	note = {Publisher: American Meteorological Soc},
	pages = {559--570},
}

@article{beucler_enforcing_2021,
	title = {Enforcing {Analytic} {Constraints} in {Neural} {Networks} {Emulating} {Physical} {Systems}},
	volume = {126},
	issn = {10797114},
	url = {https://doi.org/10.1103/PhysRevLett.126.098302},
	doi = {10.1103/PhysRevLett.126.098302},
	number = {9},
	journal = {Physical Review Letters},
	author = {Beucler, Tom and Pritchard, Michael and Rasp, Stephan and Ott, Jordan and Baldi, Pierre and Gentine, Pierre},
	year = {2021},
	pmid = {33750168},
	note = {arXiv: 1909.00912
Publisher: American Physical Society},
	pages = {98302},
}

@article{tian_double-itcz_2020,
	title = {The {Double}-{ITCZ} {Bias} in {CMIP3}, {CMIP5}, and {CMIP6} {Models} {Based} on {Annual} {Mean} {Precipitation}},
	volume = {47},
	issn = {19448007},
	doi = {10.1029/2020GL087232},
	number = {8},
	journal = {Geophysical Research Letters},
	author = {Tian, Baijun and Dong, Xinyu},
	year = {2020},
	pages = {1--11},
}

@book{francois_adjusting_2021,
	title = {Adjusting spatial dependence of climate model outputs with cycle-consistent adversarial networks},
	isbn = {0-12-345678-9},
	url = {https://doi.org/10.1007/s00382-021-05869-8},
	publisher = {Springer Berlin Heidelberg},
	author = {François, Bastien and Thao, Soulivanh and Vrac, Mathieu},
	year = {2021},
	doi = {10.1007/s00382-021-05869-8},
	note = {Publication Title: Climate Dynamics
Issue: 0123456789
ISSN: 14320894},
}

\end{document}


\maketitle


\newgeometry{textwidth=5.5in}
\tableofcontents

\section{Score-based diffusion models}

    Score-based generative diffusion models \cite{song_generative_2019, song_score-based_2021, song_denoising_2022} aim to model a reversed diffusion process in the framework of continuous-time stochastic differential equations (SDEs). The \emph{forward} diffusion process that incrementally perturbs the data can be described as the solution of the SDE:
    \begin{equation}
        \mathrm{d}\mathbf{x} =  \mu(\mathbf{x}, t) \mathrm{d}t + g(t)\mathrm{d}\mathbf{w},
         \label{eq:forward_sde_si}
    \end{equation}
    where $\mu(\mathbf{x}, t) : \mathbb{R}^d \rightarrow  \mathbb{R}^d$  is the drift term with $d$ being the dimension of an image, $\mathbf{w}$ denotes Wiener noise and $g(t) : \mathbb{R} \rightarrow  \mathbb{R}$ is the diffusion coefficient. For variance exploding (VE) SDEs \cite{song_score-based_2021}, the diffusion coefficient that acts as a variance schedule is given by
    \begin{equation}
        g(t) = \sigma_{\mathrm{min}} \left(\frac{\sigma_{\mathrm{max}}}{\sigma_{\mathrm{min}}} \right)^{t} 
        \sqrt{2 \log \left(  \frac{\sigma_{\mathrm{max}}}{\sigma_{\mathrm{min}}} \right) },
    \end{equation}
    where $\sigma_{\mathrm{min}}$ and $\sigma_{\mathrm{max}}$ are chosen s.t. $p_{\sigma_{\mathrm{min}}}(\mathbf{x}) \approx p_{\mathrm{data}}(\mathbf{x})$ and $p_{\sigma_{\mathrm{max}}}(\mathbf{x}) \approx \mathcal{N}(\mathbf{x}; \mathbf{0}, \sigma^2_{\mathrm{max}} \mathbf{I})$.
    The reverse SDE that incrementally removes noise and thus can be used to generate data is given by \cite{anderson_reverse-time_1982}
    \begin{equation}
        \mathrm{d}\mathbf{x} = [ \mu(\mathbf{x}, t) - g(t)^2 \nabla_{\mathbf{x}} \log p_{t, \mathrm{data}}(\mathbf{x}) ] \mathrm{d}\bar{t} + g(t)\mathrm{d}\mathbf{\bar{w}},
         \label{eq:reverse_sde_si}
    \end{equation}
     where $\nabla_{\mathbf{x}} \log p_{t, \mathrm{data}}(\mathbf{x})$ is the score function of the marginal data distribution. A time-dependent neural network $S(\mathbf{x}, t; \bm{\theta}): \mathbb{R}^d \rightarrow \mathbb{R}^d$ with parameters $\bm{\theta}$ is then trained with denoising score matching \cite{song_denoising_2022, vincent_connection_2011} to approximate the score function of the target distribution with
    \begin{equation}
        \frac{S(\mathbf{x},t; \bm{\theta})}{\sigma(t)} =  s(\mathbf{x},t; \bm{\theta}) \approx \nabla_{\mathbf{x}} \log p_{t, \mathrm{data}}(\mathbf{x}),
         \label{eq:score}
    \end{equation}
    where $\sigma(t)$ is given by
    \begin{equation}
        \sigma(t) = \sigma_{\mathrm{min}} \left(\frac{\sigma_{\mathrm{max}}}{\sigma_{\mathrm{min}}} \right)^{t}.
        \label{eq:sde_var}
    \end{equation}
    The loss function for the training is given by
    \begin{equation}
        \mathcal{L}(\bm{\theta}) = \mathbb{E}_{t, \mathbf{x}(0), \mathbf{x}(t)} \left[\lambda(t) ||  s(\mathbf{x},t; \bm{\theta}) - \nabla_{\mathbf{x}} \log p(\mathbf{x}(t) | \mathbf{x}(0)) | |_2^2 \right],
        \label{eq:sde_loss}
    \end{equation}
    where $\lambda(t): [0, T] \rightarrow \mathbb{R}_{>0}$ is a weighting function, $t \sim \mathcal{U}(0, T)$, $\mathbf{x}(0) \sim p_0(\mathbf{x})$, and $\mathbf{x}(t) \sim p(\mathbf{x}(t) | \mathbf{x}(0))$. 
    Since the transition kernel $p(\mathbf{x}(t) | \mathbf{x}(0))$ is given by a Gaussian \cite{song_score-based_2021} and is hence known analytically, we can compute the score $\nabla_{\mathbf{x}} \log p(\mathbf{x}(t) | \mathbf{x}(0)) = (\mathbf{x}(t) - \mathbf{x}(0))/\sigma^2$, where $\mathbf{x}(t) = \mathbf{x}(0) + \sigma(t) \epsilon$ and $\epsilon \sim \mathcal{N}(\mathbf{0}, \mathbf{I})$. The loss function in Eq.~\ref{eq:sde_loss} essentially defines a regression problem that is much more stable than the adversarial training in GANs.
    
    For our study, we set $\sigma_{\mathrm{min}} = 0.01$ and $\sigma_{\mathrm{max}} = 500$ and use a warmup of 1000 steps.

\section{Training Parameters and Dynamics}

    \begin{table}[h!]
        \small
        \centering
            \caption{\small{Summary of the training parameters for the consistency and SDE model, such as the batch size (BS) and learning rate (LR).}}
            \small{
        \begin{tabular}{r c c c c c c c}
            \toprule
            Model & BS & LR & Optimizer & EMA decay & Noise variance & Time discretization & Warmup \\
            \hline
            CM    & 1          &  $2e^{-4}$    & RADAM     &  $w_0=0.9$  & \makecell{$\sigma_{\text{data}}=0.5$,\\ $t_{\text{min}}=0.002$, \\ $t_{\text{max}}=80$} & \makecell{$s_0=2$, \\  $s_1=150$,\\ $\rho=7$} & - \\
            \hline
            SDE    & 1          &  $2e^{-4}$    & ADAM     &  $w_0=0.999$  & \makecell{$\sigma_{\text{min}}=0.01$,\\ $\sigma_{\text{max}}=500$} & - & 1000 \\
            \bottomrule
            \label{tab:training}
        \end{tabular}       
        }
    \end{table} 

    \begin{figure}[!htb]
        \centering
        \includegraphics[width=0.9\textwidth]{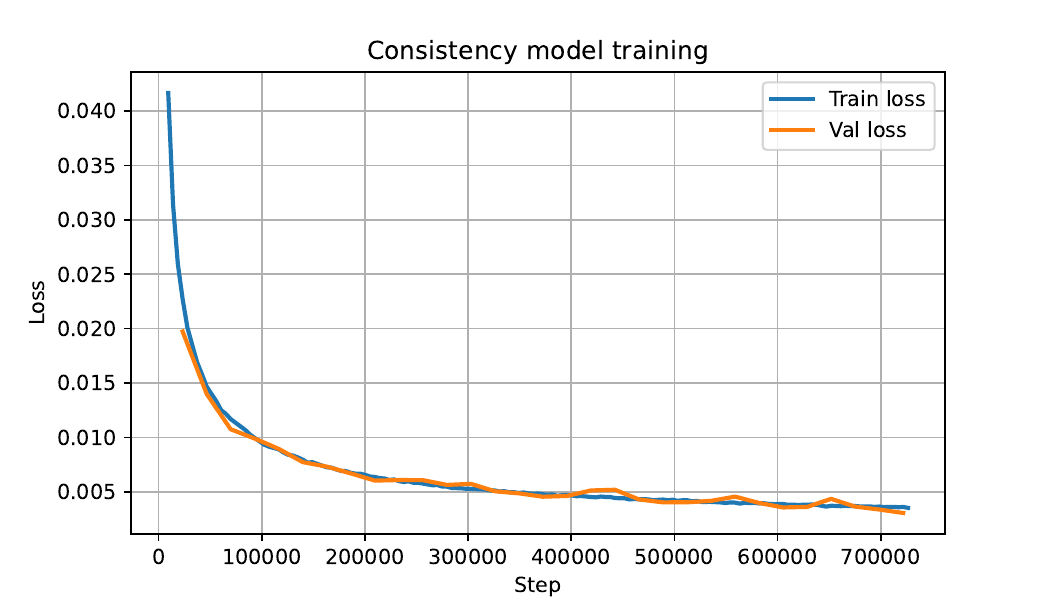}
        \caption{The training and validation loss of the consistency model is shown as a function of the update steps.}
        \label{fig:loss_dynamics}
    \end{figure}

\section{Earth System Model Overview}
An overview of the different precipitation simulations and model configurations used in this study shall be given in the following.

As a ground truth, we use ERA5 reanalysis \cite{hersbach_era5_2020}, which is the result of 4D-var data assimilation combining observations from various heterogeneous sources, such as ground weather stations and remote sensing platforms with the Integrated forecast system \cite{ecmwf_ifs_2016}, a high-resolution numerical weather prediction model. While the product is mostly simulation-based, its quality and accuracy are significantly higher than that of the numerical climate simulations used here and it has the advantage of a relatively long coverage from 1940 to the present.

The Potsdam Earth model (POEM) is used to compare the SDE and CM downscaling methods. It is a fully coupled comprehensive model including components for the atmosphere (CM2Mc) \cite{galbraith_climate_2011}, ocean (MOM5), dynamic land and vegetation LPJml5 \cite{schaphoff_lpjml4_2018, von_bloh_implementing_2018}, and ice sheets. The atmosphere has a relatively low resolution of $3^\circ \times 3.75^\circ$ and is hence relatively fast. However, the computational speed comes at a trade-off showing relatively pronounced biases, for example, in relative frequencies or long-term means. We use the model configuration of CM2Mc-LPJmL5 from \cite{druke_cm2mc-lpjml_2021} for this study.

To show that our method can be applied to more comprehensive simulations without retraining we use GFDL-ESM4 simulations \cite{dunne_gfdl_2020} from the historical CMIP6 run \cite{eyring_overview_2016}. GFDL-ESM is also a fully coupled Earth system model but has a much higher resolution and model complexity than POEM. The atmosphere of GFDL-ESM4 has a horizontal resolution of around $1^\circ$. Since we require  the resolution of the simulation to be lower than that of the ground truth, we bilinearly interpolate the GFDL-ESM4 simulation to the same $3^\circ \times 3.75^\circ$ grid in POEM to achieve the same downscaling factor.

We also include Speedyweather.jl (v0.9.0) \cite{klower_speedyweatherjl_2024}, an atmosphere-only model, in our experiments that has a lower resolution than the other atmosphere models. We use the default parameters of the wet primitive equations core, a full Clenshaw-Curtis grid with a T31 spectral truncation and 8 vertical levels for the simulations. This corresponds to a horizontal resolution of around $3.75^\circ \times 3.75^\circ$.

\section{Scale Analysis}

    \begin{figure}[!htb]
        \centering
        \includegraphics[width=0.7\textwidth]{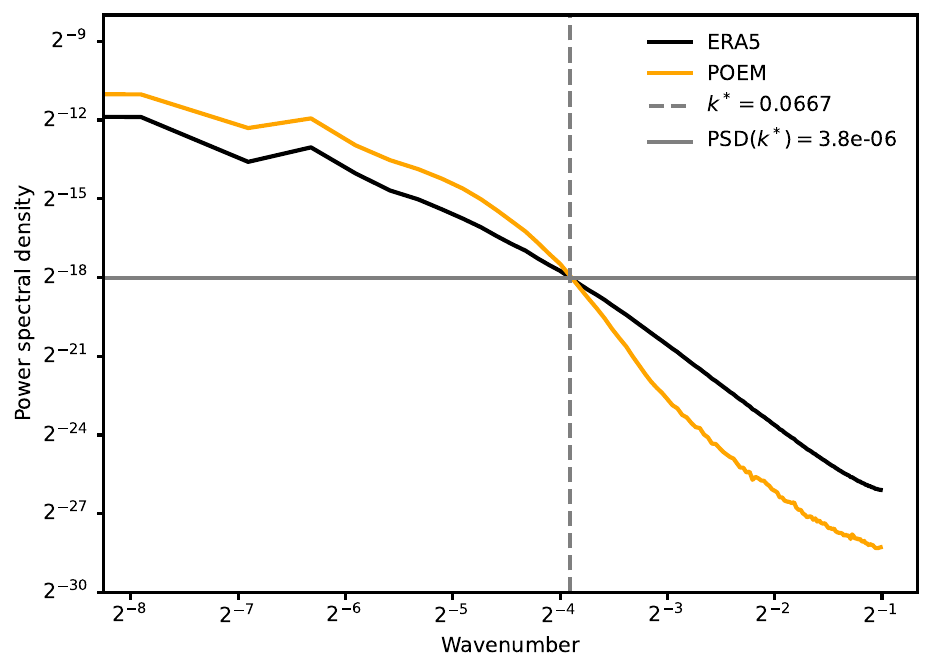}
        \caption{Power spectral densities (PSDs) of the historical ERA5 and POEM data are shown after applying the training preprocessing transformations with a log-transform and normalization to the range $[-1,1]$. The wavenumber where the PSDs of the reanalysis and ESM fields intersect is indicated at $k^*=0.0667$ as well as the corresponding PSD value of $\mathrm{PSD}(k^*)=3.8\times 10^{-6}$.}
        \label{fig:noising_scale_analysis}
    \end{figure}

\section{Downscaling Spatial Fields from Different Models}

    \begin{figure}[!htb]
        \centering
        \includegraphics[width=0.8\textwidth]{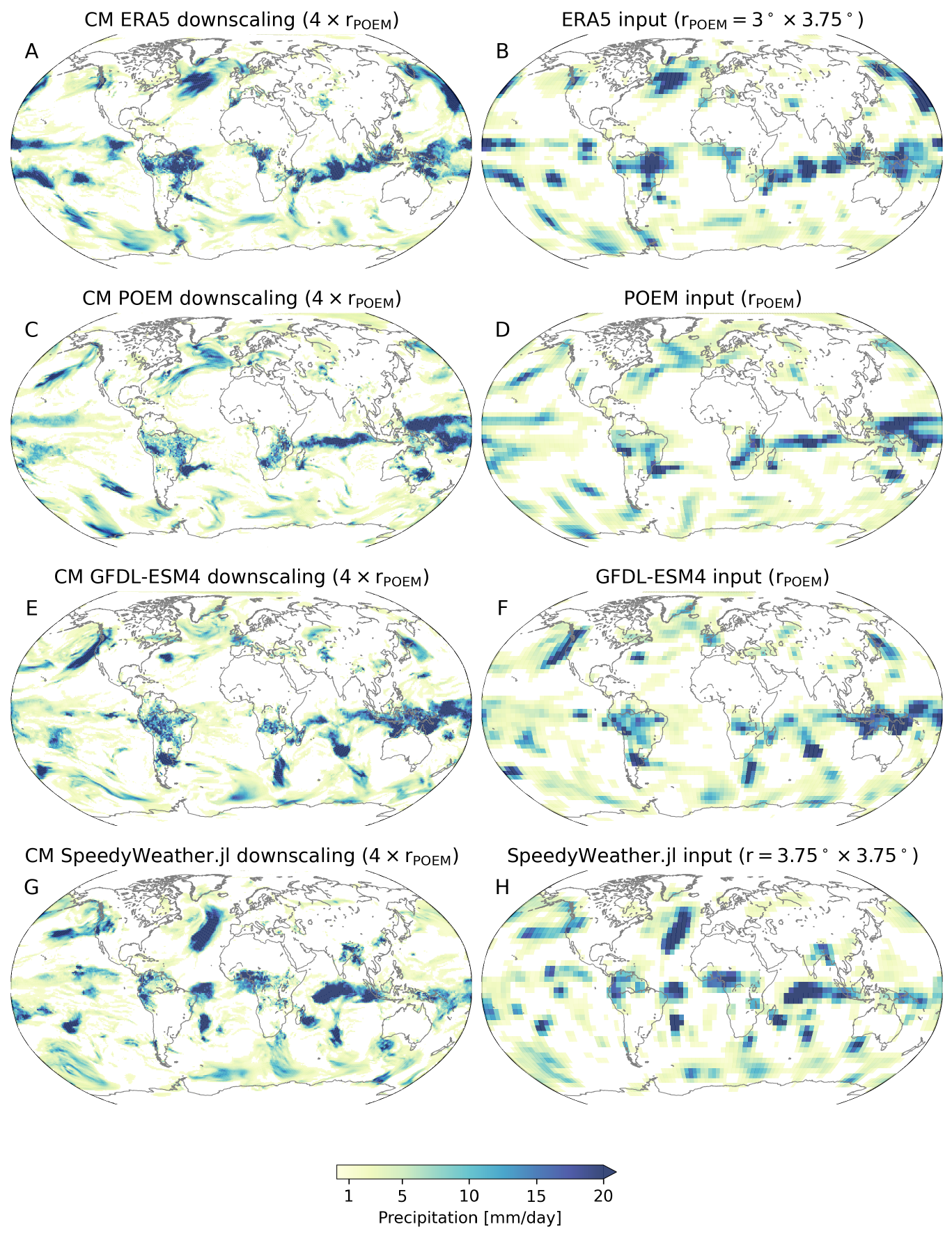}
        \caption{Qualitative comparison of the consistency model-based downscaling applied to different precipitation simulations without retraining. The left column shows the input fields and the right column the downscaled result for ERA5 (B)$\rightarrow$(A), the POEM ESM (D)$\rightarrow$(C), GFDL-ESM4 (F)$\rightarrow$(E), and SpeedyWeather.jl (H)$\rightarrow$(G).}
        \label{fig:spatial_fields_all_esms}
    \end{figure}
    
    \begin{figure}[!htb]
        \centering
        \includegraphics[width=0.9\textwidth]{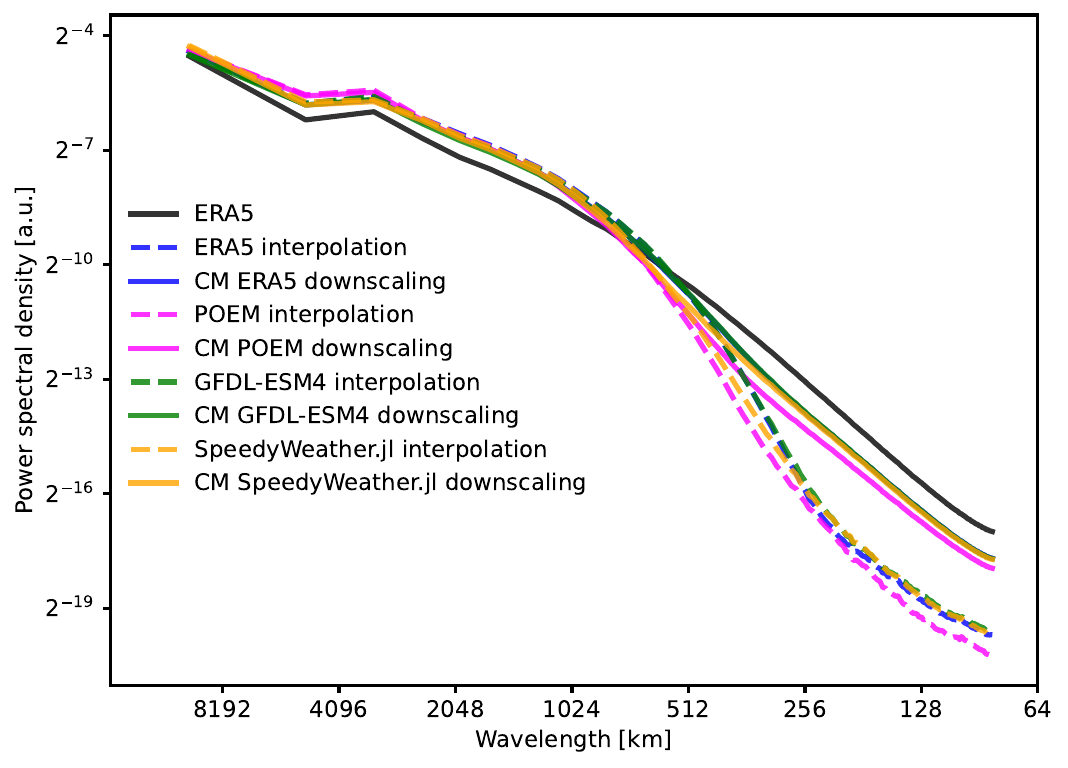}
        \caption{Averaged power spectral densities for different coarse precipitation simulations that are bilinearly interpolated (dashed lines) to the target resolution or downscaled using the consistency model (solid lines). (black) The ERA5 high-resolution ground truth, (blue) ERA5 coarsened to the POEM resolution before downscaling, (magenta) the POEM ESM, (green) the GFDL-ESM4, (orange) SpeedyWeather.jl.}
        \label{fig:pds_all_esms}
    \end{figure}

\section{Bias Correction of Different Models}

    \begin{figure}[!htb]
        \centering
        \includegraphics[width=0.9\textwidth]{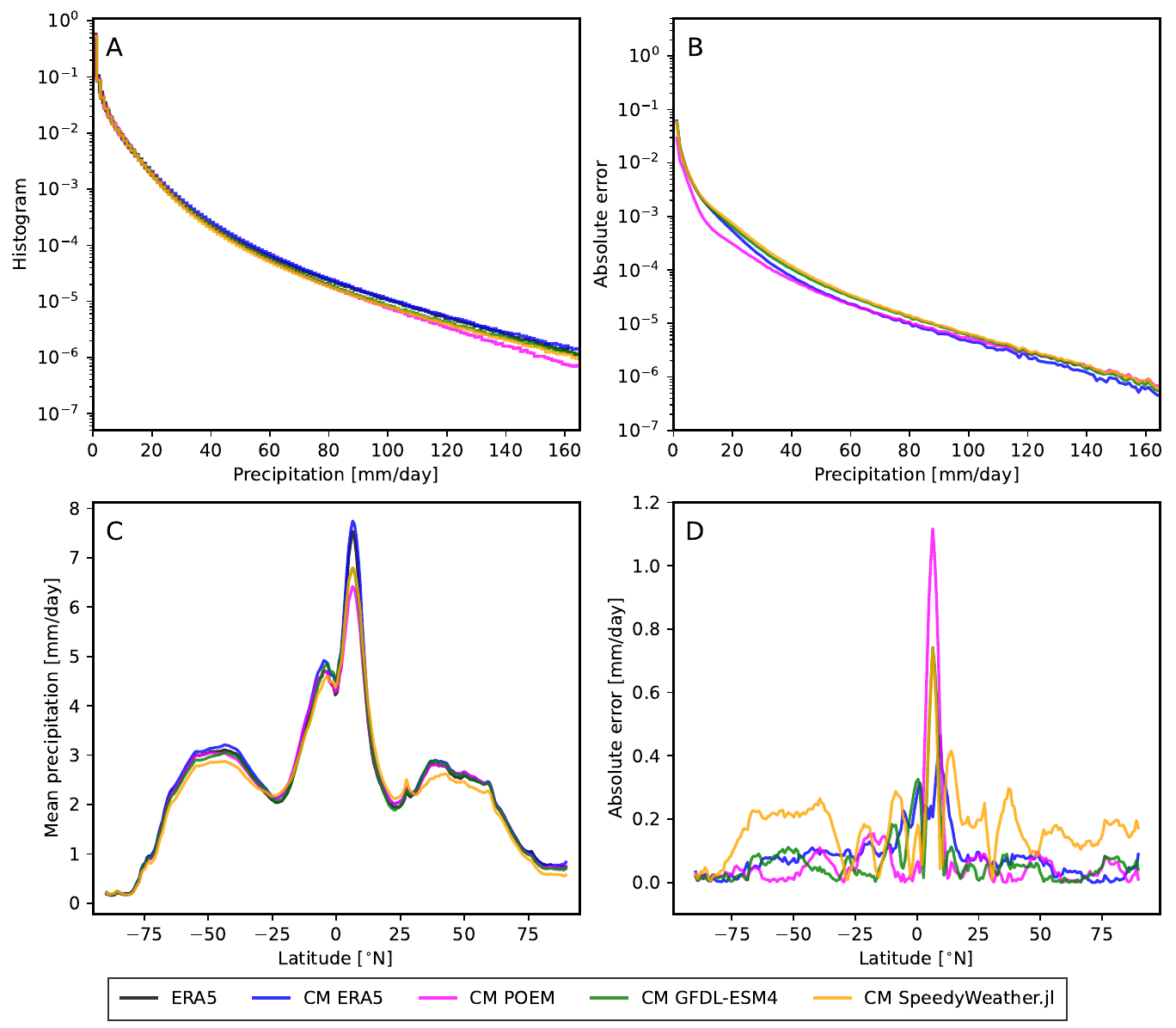}
        \caption{The bias correction is evaluated in terms of histograms and mean latitude profiles for different precipitation simulations. (black) The ERA5 high-resolution ground truth, and CM-based downscaling applied to (blue) coarse-grained ERA5, (magenta) POEM ESM, (green) GFDL-ESM4, (orange) SpeedyWeather.jl.}
        \label{fig:bias_all_esms}
    \end{figure}

\section{Downscaling ERA5 with SDE and CM}

    \begin{figure}[!htb]
        \centering
        \includegraphics[width=0.9\textwidth]{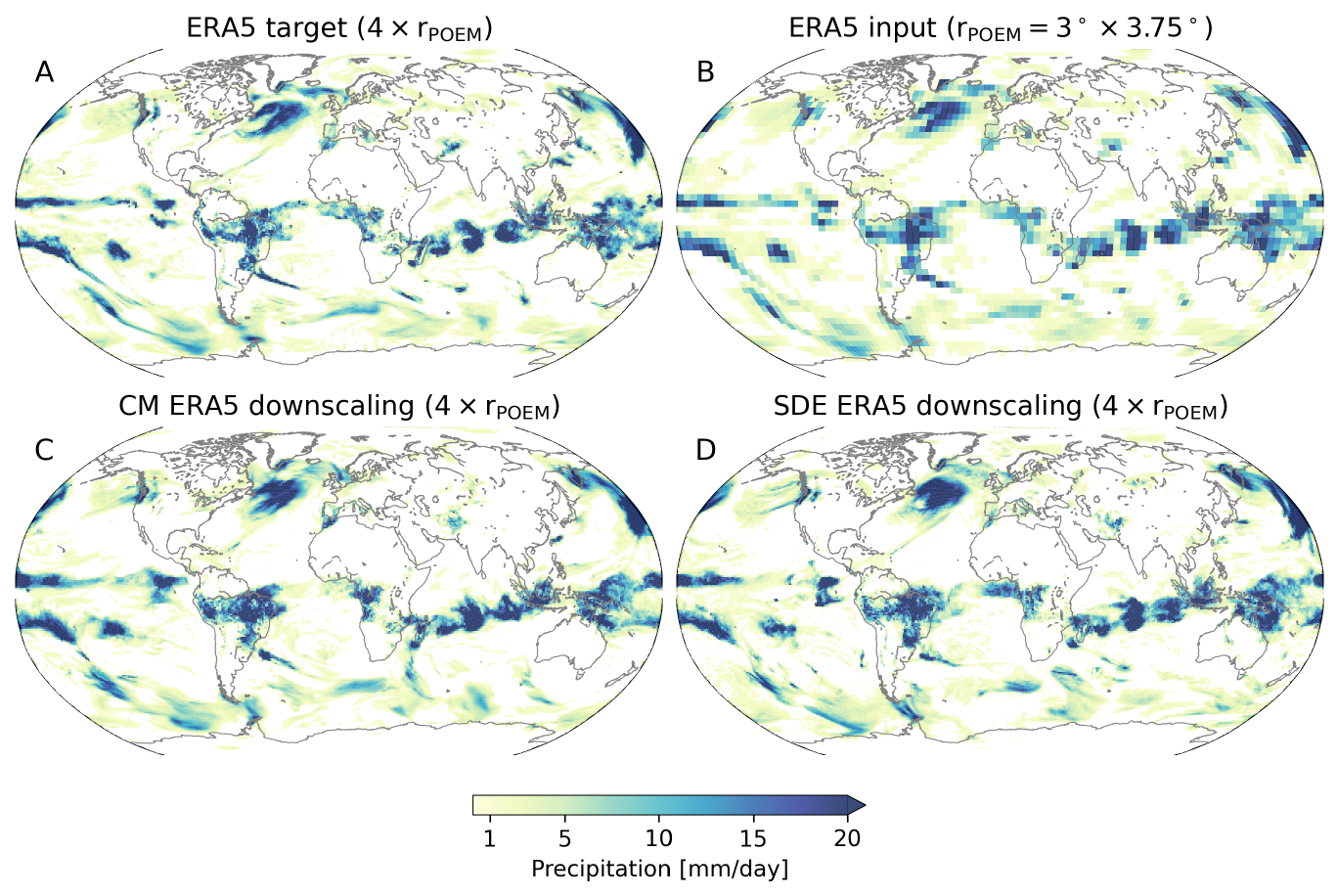}
        \caption{Qualitative comparison of downscaled precipitation fields from coarse ERA5 using the SDE and CM approaches. (A) The high-resolution ERA5 target, (B) the coarsened ERA5 used as input, (C) the CM-based downscaling, (D) the SDE-based downscaling.}
        \label{fig:spatial_fields_era5}
    \end{figure}

    \begin{table}[h!]
        \small
        \centering
            \caption{\small{Summary statistics comparing our CM approach and the SDE bridge as benchmark when downscaling ERA5 data from a resolution  of $3^\circ \times 3.75^\circ$ by a factor of four.}}
            \small{
        \begin{tabular}{r c c c c c}
            \toprule
            Model & RMSE           & MAE                & 95$^{\mathrm{th}}$ precentile error   & Corr (pooled)         & Corr (low-pass) \\
            \hline
            SDE   &  4.898         & 1.8                &  0.868                       & 0.857           & 0.902  \\
            CM    & \textbf{4.817} & \textbf{1.719}     &  \textbf{0.725}              & \textbf{0.861}  & \textbf{0.908}  \\
            \bottomrule
            \label{tab:era5_eval}
        \end{tabular}       
        }
    \end{table}

\section{Ensemble Spread Evaluation}
\label{sec:ens_spread}

    The continuous ranked probability score (CRPS) is defined as \cite{gneiting_probabilistic_2007}
    \begin{equation}
    \text{CRPS}(F, x) = \int_{-\infty}^{\infty} \left( F(y) - \mathbb{I}(y \ge x) \right)^2 \, dy,
    \end{equation}
    where $F(\cdot)$ represents the predictive distribution, in our case an ensemble of downscaled fields, $x$ is a single observation, and $\mathbb{I}(y \ge x)$ is an indicator function.
    Given the $\text{CRPS}_{\text{CM}}$ of the consistency model and a baseline $\text{CRPS}_{\text{baseline}}$, a score (CRPSS) can be defined as
    \begin{equation}
        \text{CRPSS} = 1 - \frac{\text{CRPS}_{\text{CM}}}{\text{CRPS}_{\text{CM}}}, 
    \end{equation}
    which is positive when the CM performs better than the baseline and negative for a worse performance. 
    
\begin{figure}[!htb]
        \centering
        \includegraphics[width=0.95\textwidth]{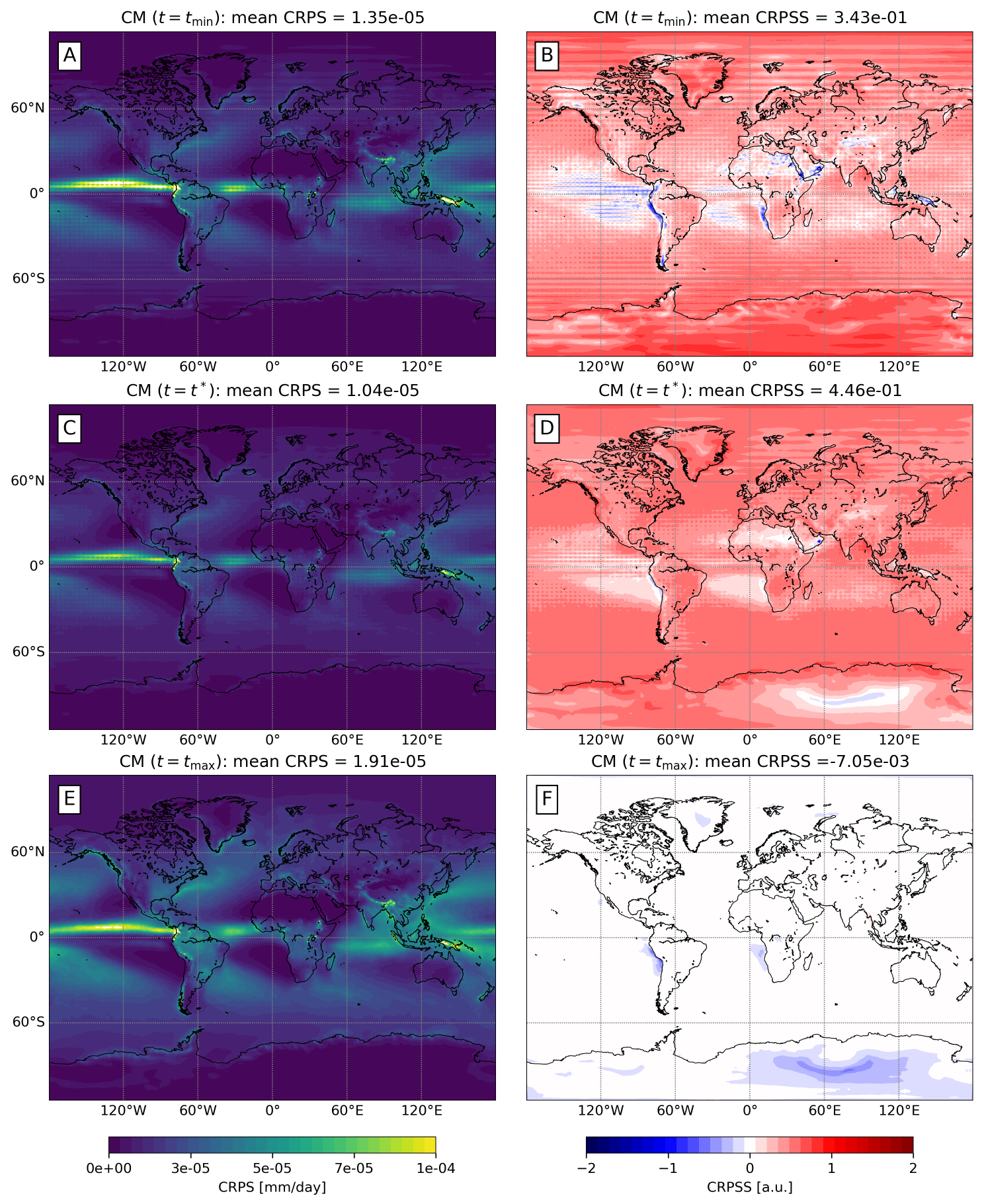}
        \caption{The continuous ranked probability score (CRPS) is computed for a generated ensemble for 100 downscaled precipitation fields for each of the coarsened ERA5 test set samples and for three different noise strengths $t_{\text{min}} < t^* < t_{\text{max}}$  in panels A, C and E. The CRP skill score (CRPSS) with respect to a random high-resolution sample baseline is shown in panels B, D, and F.}
        \label{fig:crps}
    \end{figure}

    We first downscale ERA5 fields that were initially coarsened to the POEM resolution with the CM model, creating 100 realizations for a single coarse field. We repeat this for three different noise levels corresponding to $t=t_{\text{min}}=0.002$, $t=t^*=0.63$, and $t=t_{\text{max}}=80$. As a baseline, we use 100 random samples from the high-resolution ERA5 dataset as a climatological forecast that preserves spatial correlations. The CRPS and CRPSS are then computed for the different noise levels at each grid cell individually (Fig.~\ref{fig:crps}). 
    
\section{Temporal Correlations}

    \begin{figure}[!htb]
        \centering
        \includegraphics[width=0.9\textwidth]{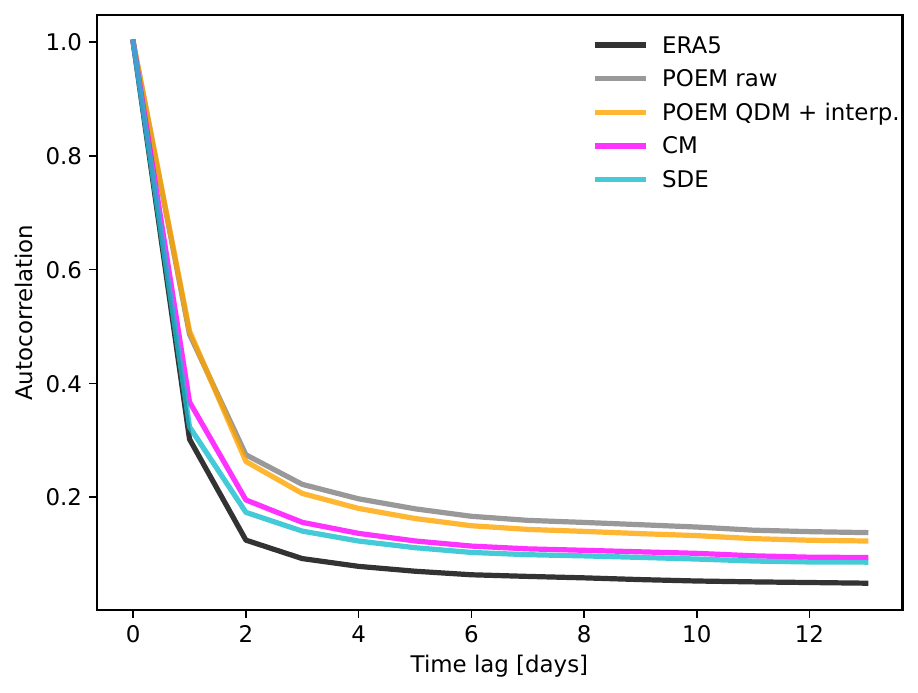}
        \caption{Temporal correlations are evaluated by computing the autocorrelation in time with a lag of up to 14 days for each grid cell. The result is then averaged globally for the high-resolution ERA5 ground truth (black), the raw POEM ESM (grey), the POEM ESM quantile mapped and interpolated to the target resolution (orange), the CM-based downscaling (magenta), and the SDE approach (cyan).}
        \label{fig:temp_corr}
    \end{figure}

\section{Biases as a Function of the Preserved Scale}

    \begin{figure}[!htb]
        \centering
        \includegraphics[width=0.9\textwidth]{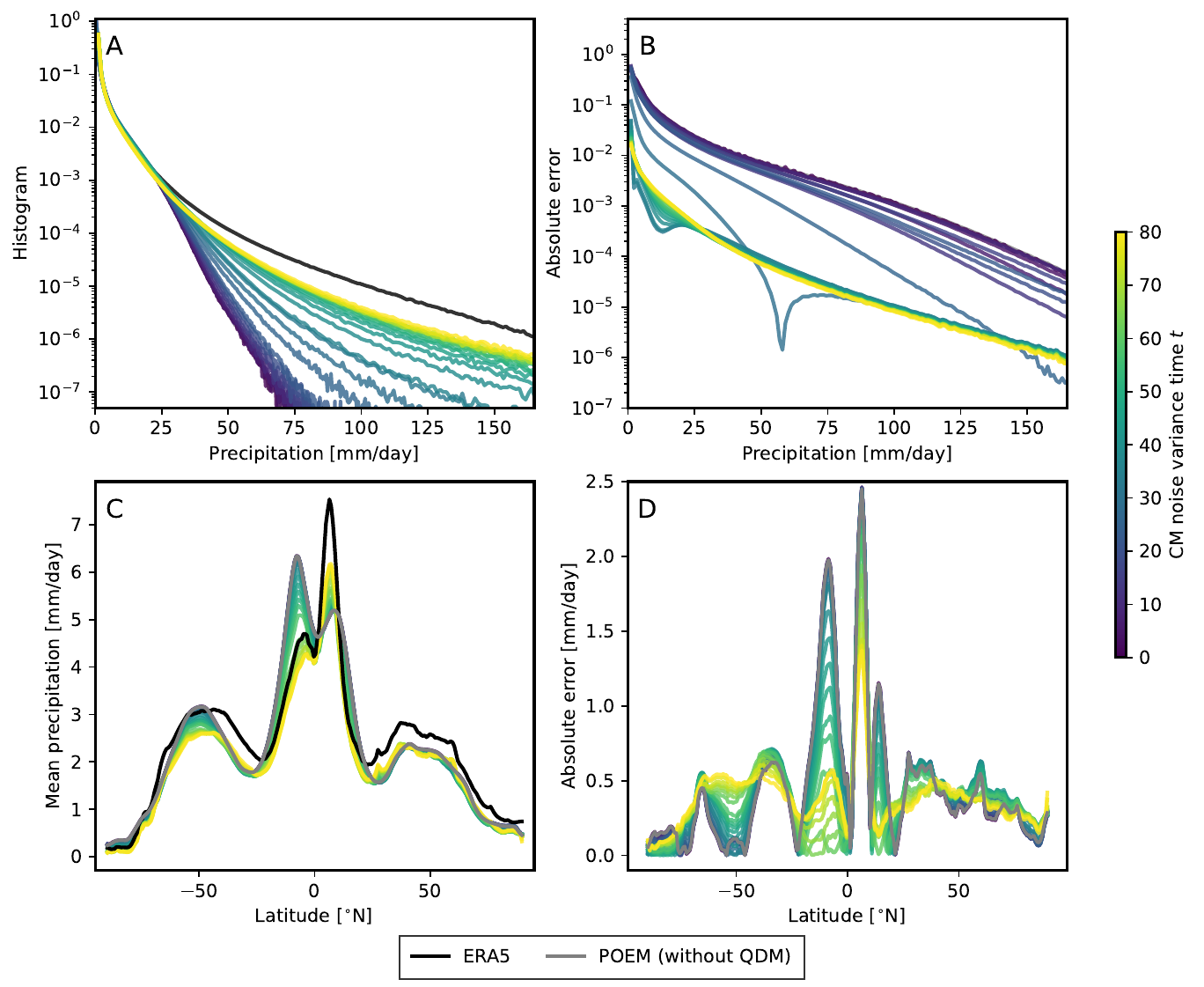}
        \caption{Biases in term of global histograms and longitude-means are shown for the ERA5 ground truth (black), the POEM ESM without QDM-preprocessing (grey) and the CM downscaling for different noise levels as a function of $t$.}
        \label{fig:bias_vs_noise_scales}
    \end{figure}

\bibliographystyle{naturemag}
\bibliography{references_zotero}  